\documentclass[preprint]{emulateapj} 
\usepackage{natbib}
\usepackage{hyperref}
\usepackage{bm,upgreek}

\newcommand{\be}{\begin{equation}}
\newcommand{\ee}{\end{equation}}

\newcommand{\rhat}{ {{\hat r}}}

\newcommand{\thestar}{\theta_\star} 
\newcommand{\qcrit}{q_0} 
\newcommand{\mdot}{{\dot M}_{{\rm disk}}}  
\newcommand{\mdisk}{M_{\rm disk}}  
\def\lta{\,\raise 0.3 ex\hbox{$ < $}\kern -0.75 em
 \lower 0.7 ex\hbox{$\sim$}\,}
\def\gta{\,\raise 0.3 ex\hbox{$ > $}\kern -0.75 em
 \lower 0.7 ex\hbox{$\sim$}\,} 
\newcommand{\teff}{ T_{\rm eff} } 
\newcommand{\gcal}{\mathcal{G}} 

\shorttitle{Disk-Star Interactions and Orbital Obliquity} 
\shortauthors{Batygin \& Adams} 

\begin{document}
 
\title{Magnetic and Gravitational Disk-Star Interactions: \\
An Interdependence of PMS Stellar Rotation Rates and Spin-Orbit Misalignments}   
\author{Konstantin Batygin$^{1}$ and Fred C. Adams$^{2,3}$} 

\affil{$^1$Institute for Theory and Computation, Harvard-Smithsonian Center for Astrophysics \\ 
60 Garden St., Cambridge, MA 02138}
\affil{$^2$Department of Physics, University of Michigan, Ann Arbor, MI 48109} 
\affil{$^3$Department of Astronomy, University of Michigan, Ann Arbor, MI 48109} 
\email{kbatygin@cfa.harvard.edu, fca@umich.edu}

\begin{abstract}
The presence of giant gaseous planets that reside in close proximity to their host stars, i.e. hot Jupiters, may be a consequence of large-scale radial migration through the proto-planetary nebulae. Within the framework of this picture, significant orbital obliquities characteristic of a substantial fraction of such planets can be attributed to external torques that perturb the natal disks out of alignment with the spin axes of their host stars. Therefore, the acquisition of orbital obliquity likely exhibits sensitive dependence on the physics of disk-star interactions. Here, we analyze the primordial excitation of spin-orbit misalignment of Sun-like stars, in light of disk-star angular momentum transfer. We begin by calculating the stellar pre-main sequence rotational evolution, accounting for spin-up due to gravitational contraction and accretion as well as spin-down due to magnetic star-disk coupling. We devote particular attention to angular momentum transfer by accretion, and show that while generally subdominant to gravitational contraction, this process is largely controlled by the morphology of the stellar magnetic field (that is, specific angular momentum accreted by stars with octupole-dominated surface fields is smaller than that accreted by dipole-dominated stars by an order of magnitude). Subsequently, we examine the secular spin-axis dynamics of disk-bearing stars, accounting for the time-evolution of stellar and disk properties and demonstrate that misalignments are preferentially excited in systems where stellar rotation is not overwhelmingly rapid. Moreover, we show that the excitation of spin-orbit misalignment occurs impulsively, through an encounter with a resonance between the stellar precession frequency and the disk-torquing frequency. Cumulatively, the model developed herein opens up a previously unexplored avenue towards understanding star-disk evolution and its consequences in a unified manner.

\end{abstract}


\section{Introduction}
\label{sec:intro} 

The birth and early evolution of planetary systems is a phenomenologically rich physical process. Among the numerous factors that affect this process is the physical evolution of pre-main-sequence (PMS) stars, around which planets form. Consequently, understanding the time-dependence of PMS stellar parameters may be crucial to understanding the primordial shaping of planetary systems.

The appreciation for the role that young stars play in sculpting planetary systems was already recognized in early investigations of solar system formation (see for example \citealt{1980PThPh..64.1968S}). However, this connection has been strengthened immensely by the discovery and subsequent characterization of extrasolar planets \citep{wolfboy,mayor,marcy,charbonneau}.  Indeed, such early results as the stellar metallicity-planet occurrence correlation
\citep{fishvalenti} have had a tremendous impact on our understanding
of planet formation \citep[e.g.,][]{laughlin2004} while from a
historical point of view, it is interesting to note that the first
proposed search (via the radial velocity and transit techniques) for
inflated giant planets that reside in close proximity to their host
stars (now known as hot Jupiters) was inspired by observations of
eclipsing binary stars \citep{struve1952}.

\begin{figure*}
\includegraphics[width=1\textwidth]{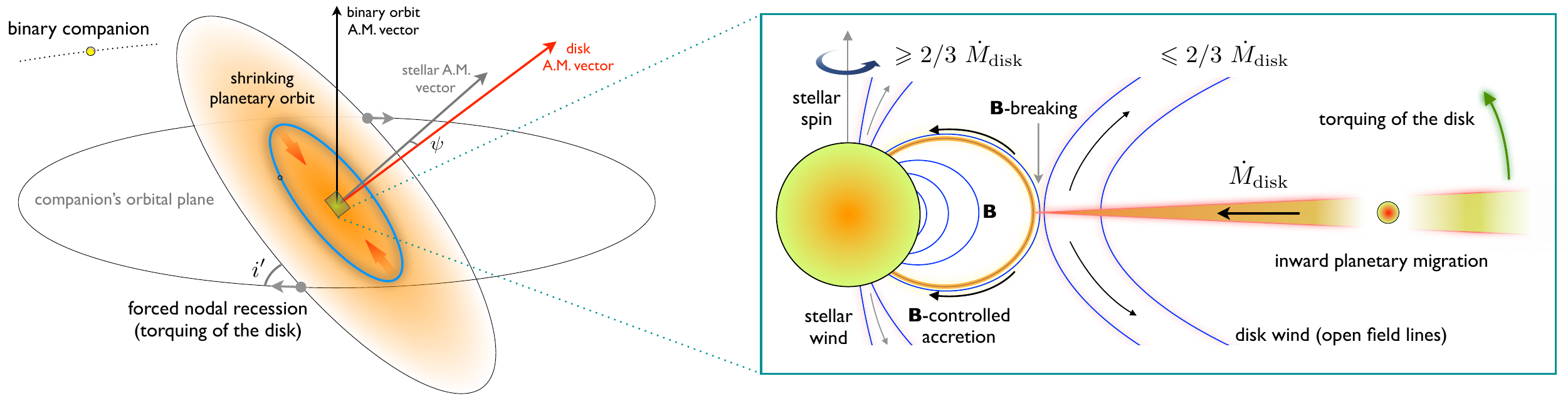}
\caption{Geometrical setup of the problem. The left side of the figure depicts a protoplanetary disk residing in a binary system, where the disk-binary orbit inclination is $i'$. The disk is taken to possess a giant planet, whose orbit slowly shrinks due to the viscous evolution of the gas (type-II migration). The angular momentum vectors of the stellar rotation, binary orbit, and the disk are labeled. Note that the (unprojected) misalignment angle $\psi$ is the angle between the angular momentum vectors of the stellar rotation and the disk. The right side of figure depicts a close-up of the disk-star interface. Blue lines depict the the stellar magnetic field, and accretion of disk material is taken to occur along a critical field line, highlighted with orange on the figure. For completeness, the figure also shows open magnetic field lines associated with the stellar and disk winds. Although disk winds remove a comparatively small fraction of the disk mass (an effect we ignore in the paper), they can carry away a significant fraction of the orbital angular momentum.} 
\label{setup}
\end{figure*}

Over the last decade or so, both stellar and planetary astronomy have
made considerable observational advances, allowing the scientific
frontiers to shift forward. With a relatively well-established
understanding of the main-sequence evolution of sun-like stars,
stellar-oriented observational surveys have began to focus on
elucidating the distribution of rotation rates \citep{herb2007} as
well as magnetic field properties \citep{gregory2012} of PMS T-Tauri stars.  Over the same time period, the
aggregate of transiting hot Jupiters has grown to an appreciable size,
and observations of the Rossiter-McLaughlin effect
\citep{rossiter,mclaugh} have revealed that the orbits of some hot
Jupiters are significantly misaligned with respect to the spin-axes of
their host stars \citep{winn2007,winn2011a,winn2011b}.

A plausible explanation for the close-in orbits of hot Jupiters is orbital migration, driven by viscous evolution of the protoplanetary disks\footnote{The same explanation is often invoked to account for the assembly of giant planets into mean motion resonances \citep{2002ApJ...567..596L,2007Icar..191..158M}.} \citep{goldreich, lin1996}. Within the context of this model, planets arrive to their short-period (of order $\sim 3$ days) orbits early on, allowing disk-star coupling to partially dictate the range of viable orbital obliquities. Accordingly, \textit{the primary purpose of this work is to consider protoplanetary disk-star interactions and propose a framework in which spin-orbit misalignments and PMS stellar evolution fit into a unified picture}. Before engaging in calculation, let us first briefly review some of the relevant observational studies. 

By the turn of this century, thousands of PMS stellar rotation rate measurements had been performed
\citep{herb2002}. This increase in available data has provided
estimates for the width of the PMS spin period distribution. Specifically, in
the observational sample of \cite{littlefair}, the rotation periods
are distributed such that most stars lie in the range $P=1-10$ days
(although some stars rotate as quickly as $P = 0.6$ days and some as
slowly as $P = 20$ days). In addition, this sample shows a bimodality
in the period distribution, with the two peaks centered around $P=2$
and $P=8$ days \citep{herb2007}. However, these studies use
ground-based observations that could be subject to uncertainty.  

A competing set of space-based observations, the COROT survey of NGC
2264 \citep{affer}, has a longer time baseline and does not find a
significant bimodal distribution (see their Figure 5). Nonetheless,
the width of the distribution is nearly the same, with most periods
falling in the range $P=1-10$ days (with extrema at $\sim$0.5 and
$\sim$19 days). Moreover, there appears to be evidence for disk-star coupling in the observational sample, as weak line T Tauri stars (those without disks) are
found to rotate somewhat faster, with a median period of 4.2 days,
compared to a median of 7.0 days for classical T Tauri stars. Furthermore, observations indicate that rotational evolution depends
strongly on stellar mass (see \citealt{bouvier2013} for a recent
review). For example, rotation periods for lower
mass stars $M_\star \lesssim 0.4 M_\odot$ are shorter than those of
higher mass stars by about a factor of $\sim 2$ on average
\citep{lamm2005}. 

To explain the observations theoretically, a number of mechanisms that
affect angular momentum evolution of the stars have been explored.
These include disk-locking, where stellar magnetic fields couple to a
nearby sector of the disk and enforce quasi-co-rotation. Although uncertainties persist \citep{bouvier2013}, it is possible that this process 
can naturally slow down the stellar rotation from breakup velocities to
periods in the range $P=1-15$ days \citep{konigl1991,shu1994}.
Additional angular momentum can be carried away by stellar wind,
which acts to further brake the rotation rate \citep{1988ApJ...333..236K, 2008ApJ...678.1109M}. Simultaneously, gravitational contraction that takes place along the PMS track acts to spin-up the stars \citep[e.g.,][]{donati2012}, as does accretion of disk material \citep{1996MNRAS.280..458A}. It is noteworthy that the aforementioned processes, with the exception of gravitational contraction, inherently depend on the detailed structure of the stellar magnetic field.

In a parallel set of developments, observations of the Zeeman effect
\citep{johns2007,donati2010} have revealed preferentially dipole and
octupole-dominated surface magnetic fields. Importantly, observations
suggest a mass and age dependence of the field morphology. As stellar age and mass increase, overall dipole field decreases slightly
while the relative strength of the octupole increases, reaching values
as much as an order of magnitude larger than those of the dipole
component \citep{gregory2012}.  Qualitatively speaking, this
dependence may be expected. More massive stars develop radiative
interiors early in their PMS evolution \citep{hansenkawal,phillips},
effectively pushing the primary field-generating region to the outer
layers of the stars. This development in turn forces the higher-order harmonics of
the field to become more pronounced at the surface. It is interesting
to note that in some similarity to this picture, the non-dipolar
dominated field of Neptune is thought to be a consequence of a stably
stratified layer in the planetary interior \citep{stanblox}.

Finally, let us remark on spin-orbit misalignments of transiting
planets. Although to date the total number of measurements continues to be
relatively small, it is believed that about a fourth of the Hot
Jupiter sample exhibits large orbital obliquities \citep{2011PASP..123..412W}. Orbital obliquities are typically interpreted as relics associated with the physical mechanism by which giant planets that form beyond the ice-line \citep{pollack1996} migrate to short-period orbits. However, \citet{2012ApJ...758L...6R} argue that the observations can be understood as a manifestation of the modulation of stellar surface rotation by internal gravity waves. Even if the conventional interpretation is correct (an assumption we shall make here), the origin of these obliquities remains elusive, as the process by which giant planets arrive onto close-in
orbits also remains controversial. Traditionally, hot Jupiters with
high orbital obliquities have been attributed to ``violent" migration
scenarios where an initially distant giant planet gets tidally
captured onto a close-in orbit after attaining a nearly parabolic
orbit through processes such as planet-planet scattering
\citep{fordrasio,nagasawa,beaugenes}, Kozai resonance
\citep{wumurray,fabtremaine,naoz}, or slow chaotic diffusion
\citep{wulithwick}. Meanwhile ``calm" disk-driven (i.e., Type-II)
migration \citep{goldreich,lin1996} was thought to produce spin-orbit
aligned planets. However, it has been recently shown that spin-orbit
misalignments naturally arise within the framework of disk-driven
migration as a result of external gravitational torques exerted on the
disks by primordial stellar companions \citep{batygin2012}. Indeed, such
external perturbations are likely if the majority of stars of interest
(i.e., solar-type stars) are born in binary or multiple systems
\citep{ghez1993,kraus2011,marks2012}. We also note that most stars are
born in embedded cluster environments \citep{ladalada} and other
stellar members can also provide external torques via stochastic
interactions \citep{adams2010,bate2010}.

Although the disk-torquing model proposed by \cite{batygin2012}
naturally avoids the necessity for specifically-molded initial
conditions inherent to violent migration models, it is sensitive to
the physics of disk-star angular momentum transfer. Intuitively, it
can be understood that if accretion of disk material is sufficiently
rapid, then the star will continuously realign with the disk and
thereby prevent the excitation of significant misalignments
\citep{lai2011}.  On the other hand, if the star rotates rapidly
enough for its gravitational quadrupole to couple adiabatically to the
disk \citep{henrardbible}, its spin pole will trail the angular
momentum vector of the disk. This means that the distributions of
stellar rotation rates and spin-orbit misalignments may be intimately
connected. In this paper, we propose that both the existence of large spin-orbit misalignments as well as perfect alignments may be understood within the framework of the disk-torquing theory in conjunction with the evolution of stellar spin and disk-star coupling during the pre-main sequence stage of the system's lifetime.

It is often the case in astrophysical problems that the input
parameters required for the calculation possess significant
uncertainties and/or the number of physical processes that need to be
treated simultaneously is large enough to limit the precision with
which any one phenomenon can be described. In this regard, the model
we aim to construct is no exception: we are simultaneously concerned
with magnetic torquing of the star by the disk, magnetically-controlled accretion of disk material, the structure and physical evolution of the star and the disk, as well as gravitational perturbations of a binary companion on the system. Indeed, each of these processes individually constitutes an active research field. Accordingly, rather than trying to capture as much detail as possible with the aid of numerical simulations, here we concentrate on developing an intuitive analytical theory, which we then use to shed light on the behavior of the system in the parameter regimes of interest.

This paper is organized as follows. In Section \ref{sec:models}, we outline the basic ingredients needed for our calculation, namely the stellar and disk structure/evolution models. In Section \ref{sec:accretion}, we consider the rotational evolution of young stars. In particular, we consider gravitational contraction, magnetic disk-star torques, as well as accretion. Specific care is taken in treating magnetically controlled accretion of disk material onto stars, as we show that (contrary to some previous claims) this process is generally sub-dominant to gravitational contraction. Moreover, geometrical considerations imply that objects with octupole-dominated surface fields accrete an order of magnitude less angular momentum than their dipole-dominated counterparts. In Section \ref{sec:dynamics}, we study the dynamics of stellar spin-axes subject to interactions with a continuously torqued disk, accounting for the evolution of the stellar rotation rate. Our calculations demonstrate that while significant misalignments cannot be excited for rapidly-rotating stars, arbitrarily large misalignments can be excited, via resonant encounters, for stars that are successfully spun-down. We discuss our results and their implications in Section \ref{sec:conclusion}.

\section{Models for Disk and Star Evolution} 
\label{sec:models} 

\begin{figure}
\includegraphics[width=1\columnwidth]{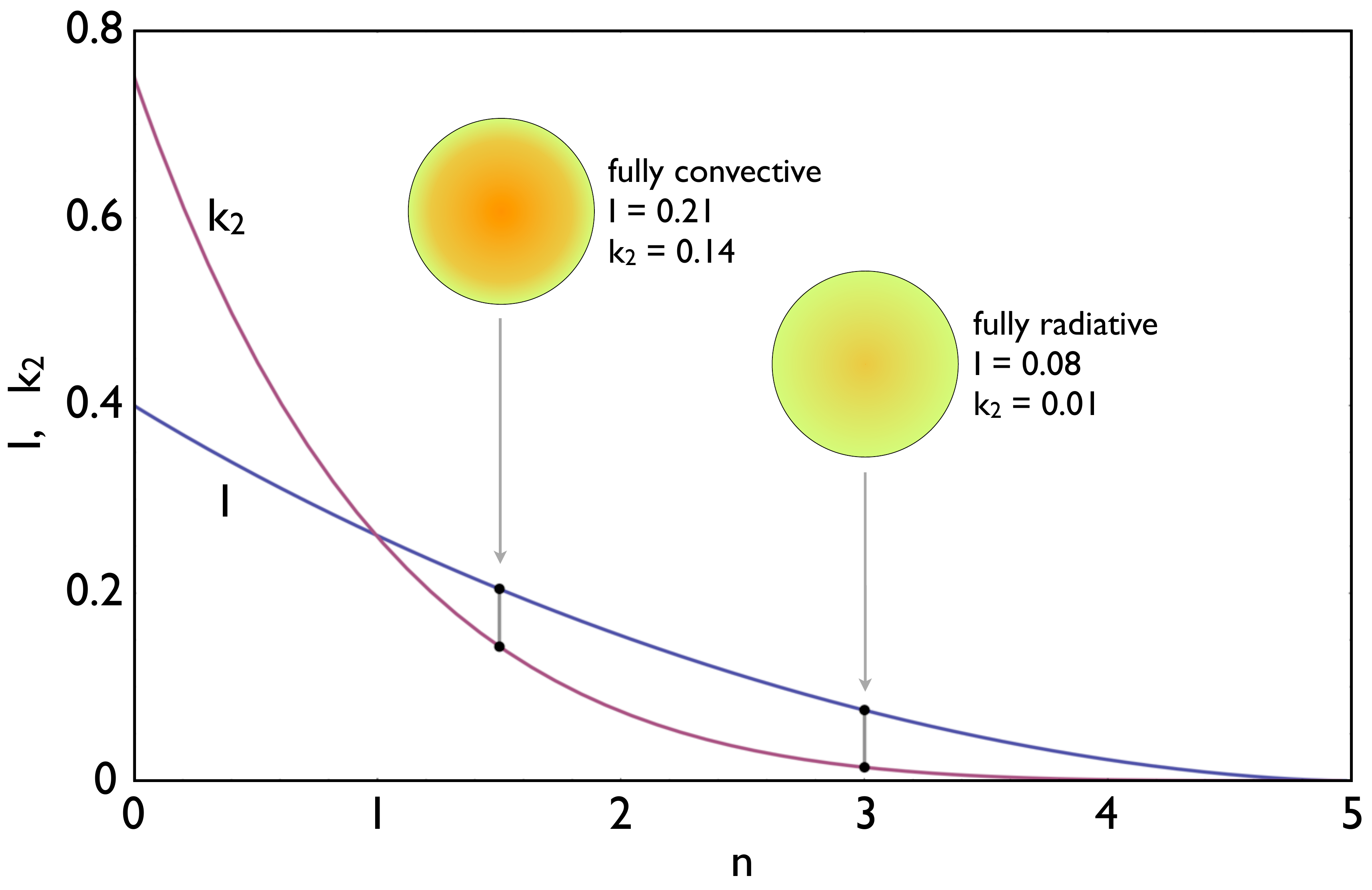}
\caption{The dimensionless moment of inertia, $I$, and the apsidal  
motion constant, $k_2$, as functions of the polytropic index, $n$. 
The $n=3/2$ polytrope, corresponding to a fully convective star and used throughout the paper, as well as a $n=3$ polytrope, corresponding to a fully radiative star (closer to the present Sun) are labeled.}
\label{Ik2}
\end{figure}

The general setup of the model we consider is depicted as a cartoon in
Figure \ref{setup}. As mentioned above, we shall begin by treating
the evolution of the stellar spin, accounting for star-disk angular momentum transfer. To complete the specification of
the problem we need to determine the basic parameters for the star and
disk. First, we need to specify the mass accretion rate $\mdot$ and
the disk mass $\mdisk$ as functions of time. In the simplest case of
isolated disks, possessing no additional sources or sinks of disk
material, the two quantities are related via $d\mdisk/dt = \mdot = -
\dot{M}_{\star}$. Here, we adopt this simplifying assumption. For
completeness, however, we note that the disk can lose mass through
photoevaporation, in addition to accreting mass onto the star.
Further, in the embedded phase of evolution, the disk can gain mass
from the infalling envelope. Thus, in a more sophisticated iteration
of our model, the two quantities $\mdot$ and $\mdisk$ could be
specified independently.

For the sake of definiteness, we use a simple evolutionary model where
the disk mass varies with time according to the function 
\be
\mdisk = \frac{\mdisk^0}{1 + t/\tau}. 
\ee 
The corresponding mass accretion rate is then given by 
\be
\mdot = -{\mdisk^0 \over \tau (1 + t/\tau)^2}\,,
\ee
where we adopt a sign convention so that the accretion rate onto the
star positive\footnote{Note that in this treatment, the increase in 
stellar mass associated with accretion is neglected.}. This general 
form produces disk masses and disk accretion rates that are roughly
consistent with observations
\citep[e.g.,][]{hart1998,hart2006,hart2008,herczeg2008,lynne2008}.  
A reasonable fit to the observed correlations can be obtained using a
starting disk mass $\mdisk^0 = 0.05 M_\star$ and a time scale $\tau_0$
= 0.5 Myr. This form can also reproduce the observed correlation between accretion rate and stellar
mass, $\mdot \propto M_\star^2$ \citep{lynne}, if we allow the time scale to vary
according to the relation $\tau = \tau_0 (M_\star/1 M_\odot)^{-1}$.
However, we note that in practice the accretion of disk material onto
young stellar objects is expected to be episodic (e.g., see
\citealt{garatti} and references therein), so the above formulae
are intended as approximate, time-averaged representations.

To specify the stellar properties, we utilize traditional polytropic
models \citep{chandra}.  After solving the Lane-Emden equation to find
the density profile, we determine the dimensionless moment of inertia
$I$ and the apsidal motion constant $k_2$ as a function of the
polytropic index $n$. Both of these quantities are required for the
calculation of stellar spin-up, as well as the spin-axis dynamics, and
are shown as functions of the polytropic index in Figure \ref{Ik2}.
Since Sun-like stars (which are the primary focus of this work) are expected to remain (partially) convective throughout much of their disk-bearing lifetime, we can use a polytopic index $n=3/2$ to model the interior structure. We note that stars that retain their disks substantially longer than $\sim 4$ Myr will develop significant radiative cores, rendering $n=3$ a more appropriate polytopic index at these latter epochs. Here, we shall ignore this complication, but for reference, both polytopic indexes are labeled in Figure \ref{Ik2}.

It is implicitly understood that the radii (and more generally the
interior structures) of stars can evolve considerably on the PMS
track.  In an approximate sense, this evolution can be understood by
noting that for a polytrope of index $n$, the energy can be written in
the form 
\be 
E = - {3 \over 10 - 2n} {\mathcal{G} M_\star^2 \over R_\star} 
= - b {\mathcal{G} M_\star^2 \over R_\star} \, ,
\ee
where the second equality defines the dimensionless parameter $b$. 
To leading order, we can assume that the energy is radiated away 
according to
\be
{dE \over dt} = - L_\star = - 4 \pi R_\star^2 \sigma T_{\rm{eff}}^4 = 
b {\mathcal{G} M_\star^2 \over R_\star^2} {d R_\star \over dt} \, . 
\label{drdt}
\ee
where the temperature is kept constant (this assumption is strictly
true only as long as the star remains convective). Neglecting changes
in the interior distribution of mass (that is, keeping $b$ = 
{\sl constant}), this expression can be considered as a differential
equation for the stellar radius, and has the solution 
\be
R_\star = \left[ {b \mathcal{G} M_\star^2 \over b G M_\star^2 / (R_\star^0)^3
+ 12 \pi \sigma T_{\rm{eff}}^4 t} \right]^{1/3} \,,
\label{roft}
\ee
where $R_\star^0$ is the stellar radius at some (arbitrarily chosen) initial condition $t=0$.

\section{Evolution of Stellar Rotation Rates} 
\label{sec:accretion} 

With stellar and disk structure models in place, this section considers the evolution of stellar spin rate due to the physical changes in the stellar radius and interactions with the disk. The primary aim of the following calculations is to identify the dominant effects and establish the conditions under which the observed period distribution can be (even approximately) reproduced. As such, we shall treat each effect in isolation and opt to perform numerical evolutionary calculations only at the end. 

Although we only seek to describe stellar rotational evolution during the PMS stage, we must further limit the scope of our treatment to obtain a sound qualitative understanding. Specifically, here we shall focus on evolution that takes place \textit{after} the embedded phase of the star-disk system. In terms of absolute stellar age, the embedded phase lasts $\sim 0.1 - 0.5$ Myr on average \citep{2009ApJS..181..321E}. As such, here we shall take the fiducial $t=0$ starting condition as the time when the properties of the young star-disk system first become optically visible. We begin by considering the most physically straight-forward process, that is gravitational contraction. 

\subsection{Gravitational Contraction}

As already alluded to above, stellar radii undergo drastic changes in the first few Myr of their lives. At birth, the radius of a Sun-like star can exceed that of the present Sun by as much as a factor of $\sim 5$, although by the time such an object emerges from the embedded phase, its radius is down to $R_\star^0 \simeq 4 R_{\odot}$ \citep{siess}. Subsequent contraction of a solar-mass star is well approximated by equation (\ref{roft}) with $T_{\rm{eff}} = 4000$K. 

In isolation, gravitational contraction conserves angular momentum. The stellar spin-up associated with the reduction of the moment of inertia is governed by the equation
\begin{equation}
2 I M_\star R_\star \omega_\star \frac{d R_\star}{dt} + I M_\star R_\star^2 \frac{d \omega_\star}{dt} = 0.
\label{notorqu}
\end{equation}
Upon substituting equation (\ref{drdt}) into this expression, it is apparent that the spin-up timescale associated with gravitational contraction is
\be
\tau_{R_\star} = {b\gcal M_\star^2 \over 8\pi\sigma T_{\rm{eff}}^4 (R_\star^0)^3} \sim 1 \rm{Myr},
\ee
which is essentially the Kelvin-Helmholz time. This timescale varies substantially with stellar mass and other properties. For instance, M dwarfs are characterized by considerably larger values of $\tau_{R_\star}$ than Sun-like stars (see \citealt{stahler} for further discussion). 

Augmented with equation (\ref{roft}), equation (\ref{notorqu}) admits the solution
\begin{equation}
\omega_\star = \omega_\star^0 \left(1+\frac{12 \pi (R_\star^0)^3 \sigma T_{\rm{eff}}^4}{b GM_\star^2}t \right)^{2/3},
\label{gravspinup}
\end{equation}
where $\omega_\star^0$ is the initial condition. Substituting the stellar parameters adopted above, as well as a typical disk lifetime of $\tau_{\rm{disk}} \sim 3$Myr \citep{2011ARA&A..49...67W} into the above expression, we obtain a factor of $\sim 4$ spin-up during the disk-bearing phase of the star. As will be shown below, the increase in the rotation rate associated with gravitational contraction dominates over that associated with accretion of disk material.

\subsection{Magnetically Controlled Accretion}

In orbit around an unmagnetized object, an accretion disk can be envisioned to extend all the way down to the object's surface\footnote{In this case, angular momentum transport is accomplished primarily by (magneto)sonic waves excited in the boundary layer \citep{1987MNRAS.228....1N, 2013ApJ...770...68B}.}. This picture no longer holds if the central star possesses a magnetic field, whose pressure can overcome the ram pressure of the accreting material \citep{ghoshlamb}. Indeed, the $\sim$ kGauss fields generally inferred for T-Tauri stars are almost certainly strong enough to carve out a substantial cavity within the inner disk \citep{shu1994}. 

The truncation radius of the disk is given by the expression
\be
a_{\rm in} = \bar{c} 
\left(\frac{8 \pi^2}{\mu_0^2} {{\cal M}^4 \over \gcal M_\star \mdot^2} \right)^{1/7}\,,
\ee
where $\bar{c} $ is a dimensionless constant of order unity and $\cal M$ is the stellar magnetic moment. Practically, irrespective of the surface morphology of the stellar field, the magnetic moment can be approximated by the dipole component alone (i.e. $\cal M = B_{\rm{dip}} R_{\star}^3/2$), since higher-order harmonics of the field decay rapidly with radius. For typical young stars, $a_{\rm{in}}$ evaluates to values not too different from the coronation radius, $a_{\rm{co}} = (G M_{\star}/\omega_\star^2)^{1/3}$. A particular state where the two are set identically equal to each other is referred to as a disk-locked condition \citep{subu}.

For a given mass accretion rate ${\dot M}$, the amount of added angular momentum depends not only on the launching point, $a_{\rm{in}}$ (in this context, the launching radius refers to the point where the accretion flow climbs out of the disk plane to form the accretion column, not to be confused with similar terminology referring to outflows), but also on the
structure of the magnetic fields. This dependence arises because in
the limit where the magnetic field is strong enough for the associated
Lorentz force to be dominant, the system takes on a
magnetohydrodynamic force-free state, where the cross product of the
fluid velocity and magnetic field vanishes \citep[e.g.,][]{moffatt}.
In other words, the accretion flow follows magnetic field lines, so 
that the field geometry determines, in part, the amount of angular 
momentum added to the star. 

For a given magnetic field configuration, we can construct coordinate
systems where one coordinate follows the fields (and hence the flow),
while the other two coordinates are orthogonal (following
\citealt{ag2012}; see also \citealt{adams2011}). For tractability we
restrict the flow to the poloidal plane, which leads to axisymmetric
flow. With this simplifaction, we can choose the usual azimuthal angle
$\varphi$ as the third coordinate. The remaining two coordinates
$(p,q)$ depend on the magnetic field geometry. The gradient
$\nabla{p}$ points in the direction of the magnetic field vector,
whereas $\nabla{q}$ is perpendicular so that $\nabla{p}\cdot\nabla{q}$
= 0. We are assuming that the magnetic field is current-free and
curl-free in the region between the inner disk edge and the stellar
surface where we are considering the flow\footnote{Currents exist
  within the star and most likely within the disk, and act as source
  terms for the magnetic field. However, strictly speaking, our
  approximation does not even require the absence of currents in the
  exterior regions, only that the magnetic field geometry can be
  written in the forms indicated.}.

To start, we take the stellar magnetic field to have both dipole and
octupole components, so that the field has the form
\begin{eqnarray}
{\bf B} &=& {B_{\rm oct} \over 2\xi^5} 
\left[ \left(5 \cos^2\theta - 3 \right) \cos\theta \,{\bf\rhat} + 
{3 \over 4} \left(5 \cos^2\theta - 1 \right) \sin\theta \, \hat{\bm\uptheta} \right] 
\nonumber \\
&+& {B_{\rm dip} \over 2\xi^3} 
\left( 2 \cos \theta \, {\bf\rhat} + \sin\theta \, \hat{\bm\uptheta} \right) \, , 
\label{dipoctfield} 
\end{eqnarray} 
where $\xi=r/R_\star$ is the dimensionless radius and $\theta$ is the
polar angle in a spherical coordinate system centered on the star.
For both the dipole and octupole terms, the leading factors of 1/2 are
included so that $B_{\rm dip}$ and $B_{\rm oct}$ are the polar
strengths of the dipole and octupole components (see also
\citealt{gregory2010,ag2012}).  It is convenient to scale out the
dipole field strength, so that the relative size of the octupole
contribution is given by the dimensionless parameter
\be 
\Gamma \equiv {B_{\rm oct} \over B_{\rm dip}} \, .  
\ee  
For T Tauri star-disk systems, observations of magnetic accretion
signatures indicate that this parameter typically falls within the
range $0\le\Gamma\le10$ (e.g., see 
\citealt{donati2007,donati2008,donati2010,donati2012}, and also 
\citealt{gregory2010}).  

For the sake of definiteness, this paper assumes that the dipole and
octupole moments are parallel. In some observed T Tauri systems,
however, the magnetic field configurations have octupole moments that
are nearly anti-parallel with the dipole moment (i.e., the main
positive pole of the octupole is coincident with the negative pole of
the dipole). To account for such systems, one can replace $\Gamma$
with $-\Gamma$ (see also \citealt{gregdon2011}), although we leave a
full exploration of this issue for future work. For completeness, we
also note that the strengths of both components $B_{\rm dip}$ and
$B_{\rm oct}$, and hence their ratio $\Gamma$, are observed to vary
with time for individual stars; as one example, the value of $\Gamma$
for the T Tauri system V2129 Oph was observed to vary by a factor of 2
over the four year period 2005 -- 2009 \citep{donati2011}.  Although
we expect such variations on timescales of $\sim$ one year, over longer
timescales the ratio $\Gamma$ does appear to increase with stellar age
when all the stars with dipole-octupole fields are considered.

With the magnetic field configuration of equation (\ref{dipoctfield}),
the scalar fields that define the coordinate system take the form 
\begin{eqnarray}
\mathbf{p} &=& - {1 \over 4} \xi^{-4} \Gamma \left( 5 \cos^2\theta - 3 \right) 
\cos\theta - \xi^{-2} \cos\theta, \nonumber \\
\mathbf{q} &=& {1 \over 4} \xi^{-3} \Gamma \left( 5 \cos^2\theta - 1 \right) 
\sin^2\theta + \xi^{-1} \sin^2\theta \, . 
\label{qdef} 
\end{eqnarray} 
Note that the coordinates are dimensionless in this treatment. 
Notice also that for a purely dipole field, $\Gamma \to 0$, and the 
coordinates $(p,q)$ simplify to the form (see also \citealt{radoski}): 
\be
\mathbf{p} = - \xi^{-2} \cos \theta \qquad {\rm and} \qquad 
\mathbf{q} = \xi^{-1} \sin^2 \theta \, . 
\label{dipoledef} 
\ee 

Let us now consider the field lines that connect the inner disk edge
to the stellar surface. To a good approximation, it is sufficient to
consider only a single ``critical'' field line that intersects the
fiducial truncation radius of the disk (\citealt{ghoshlamb}; see also
\citealt{subu}).  At the location of the inner edge of the disk,
$a_{\rm{in}}$, the angle $\theta$ = $\pi/2$, and $\xi$ =
$\xi_{\rm{in}}$ = $a_{\rm{in}}/R_\star$. The value of the coordinate
$\qcrit$ that labels the critical magnetic field line that connects
the inner disk edge to the star is thus given by 
\be
\qcrit = {1 \over \xi_{\rm{in}}} 
\left( 1 - {\Gamma \over 4 \xi_{\rm{in}}^2} \right) \,.  
\label{qconnect} 
\ee
In the dipole limit, $\Gamma\to0$ and $\qcrit \to 1/\xi_{\rm{in}}$.
With the value of $\qcrit$ set by equation (\ref{qconnect}), we can
evaluate the coordinate at the stellar surface to find the polar angle
$\thestar$ where the magnetic field line (streamline) intercepts the
star (where $\xi$ = 1). We thus obtain 
\be
\qcrit = (\Gamma + 1) \sin^2\thestar - 
{5 \Gamma \over 4} \sin^4 \thestar \, , 
\label{solutone} 
\ee 
which has solution 
\be 
\sin^2 \thestar = {2 \over 5 \Gamma} \left\{ (1 + \Gamma) 
\pm \left[ (1 + \Gamma)^2 - 5 \qcrit \Gamma \right]^{1/2} \right\} \, .  
\label{soluttwo} 
\ee 
For most cases of interest, the dipole component of the field
dominates at the disk truncation radius, and we must take the minus
sign in the above expression. Furthermore, we are interested in cases
where $\qcrit$ is small, so the leading order expression becomes 
\be 
\sin^2 \thestar \approx {\qcrit \over 1 + \Gamma} \approx 
{1 \over \xi_{\rm{in}} (1 + \Gamma) } \,.
\label{sinestar} 
\ee 

With the above results in hand, the specific angular momentum $j$ of a
parcel of gas that is accreted along the critical field line is given
by the expression 
\be
j = n_{\rm{in}} R_\star^2 \sin^2 \thestar \approx 
{n_{\rm{in}} R_\star^2 \over \xi_{\rm{in}} (1 + \Gamma)} = 
{n_{\rm{in}} R_\star^3 \over a_{\rm{in}} (1 + \Gamma)} \,,
\label{jspecific} 
\ee
where $n_{\rm{in}} = \sqrt{G M_{\star}/a_{\rm{in}}^3}$ is the (nearly
Keplerian) orbital frequency (i.e., mean motion) at the disk
truncation radius $a_{\rm{in}}$. Due to the rapid fall-off of the
octupole component of the field with radius, $n_{\rm{in}}$ is nearly
the same for systems with pure dipole fields and those with
substantial octupole components, and corresponds to the rotation
frequency $\omega_\star$ of the host star under the disk-locked condition.  Thus, as stars increase
their octupole component (and hence $\Gamma$), the corresponding change
in specific angular momentum of the accreted material is produced
primarily by changes in $\thestar$\footnote{We note that the amount of
 mass accreted by the star is not necessarily affected by the changes
  in the accretion rate of angular momentum. This mismatch occurs
  because a small fraction of the disk's mass is carried away by disk
  winds (see Figure \ref{setup}), which also carry away the angular
  momentum not accreted by the star.}.

Along this line of reasoning, the work of \cite{shu1994} emphasizes
that regardless of the field morphology, the specific angular momentum
added to the star is always smaller than that at the launching point
in the disk (contrary to what was originally envisioned by
\citealt{ghoshlamb} in the context of neutron stars). Indeed, while disk winds are responsible for carrying away a relatively small fraction of the mass, they carry away the dominant fraction of angular momentum. 
At the same time, however, it is important to understand that the specific angular
momentum of the accreted material can still be larger than that of the
star. This ordering can be seen by explicitly writing out the ratio of
the angular momenta, i.e., 
\be
{j_{\rm{in}} \over j_{\star}} = 
{R_\star \over (1+\Gamma) I a_{\rm{in}}} \,,
\label{jratio}
\ee
where $I$ is the dimenionless momentum of inertia (recall that
$I\sim0.2$ for fully convective stars). As a result, this ratio is
often close to unity, so that the accreted material --- when acting in
isolation --- can either spin up or spin down the star.

The presence of $\Gamma$ in the denominators of equations
(\ref{jspecific}) and (\ref{jratio}) plays an important role. Recall
that quantitiatively, as certain stars evolve and develop stronger
octupole components, the factor $\Gamma$ increases from $\sim0$ to
$\sim10$. As a result, since $j\propto1/(1+\Gamma)$, 
{\it the specific angular momentum of the accreted material}
{\it decreases by an order of magnitude}. 

The amount of accreted angular momentum depends sensitively on the
launching radius $a_{\rm in}$. As already mentioned above, for typically observed mass accretion
rates, where $\mdot \sim 10^{-8} M_\odot$ yr$^{-1}$, the
truncation radius lies in the range $a_{\rm in} \sim 5-15 R_\star$.
However, disk accretion is known to be highly episodic. If most of the
accretion takes place during the brief time intervals when the
accretion rates are high, then the value of $a_{\rm in}$ that one
should use in the present context, for determining the amount of
angular momentum added due to disk accretion, should be that of the
outburst state. 

During the outburst phase, the
accretion rate can surge up to $\mdot\sim10^{-6} M_\odot$ yr$^{-1}$
\citep{garatti}; for extreme cases, such as FU Ori itself, the mass
accretion rate can reach $\mdot\sim10^{-5} M_\odot$ yr$^{-1}$. As a
result, the truncation radius can move inward by factors of $\sim4-7$,
so that $a_{\rm{in}} \sim 1.25-2.5 R_\star$.  Accordingly, if the
truncation radius falls in this range when most of the angular momentum
is accreted by the star, then the ratio in equation (\ref{jratio})
varies over the range $(2-4)/(1+\Gamma)$. Thus, a notable
amount of angular momentum can be added to the star if the field
configuration is dominated by the dipole component, but little angular
momentum is added if the octupole dominates.

The spin-up of a star forced by the combined effects of gravitational contraction and accretion of angular momentum is governed by the following differential equation:
\begin{eqnarray}
&2& I M_\star R_\star \omega_\star \frac{d R_\star}{dt} + I M_\star R_\star^2 \frac{d \omega_\star}{dt} \nonumber \\
 &=& {\omega_\star \over (1+\Gamma)} {R_\star^3 \over \langle a_{\rm in}\rangle} {\mdisk^0 \over \tau} {1 \over (1+t/\tau)^2}.
\label{gravaccretion}
\end{eqnarray}
It is noteworthy that the above expression is approximate
because it uses the smooth, time-averaged mass accretion rate, but
assumes that the material is accreted from a closer distance, given by
the mean value $\langle a_{\rm in} \rangle$, which is characteristic
of the launching radius for accretion during the outburst state. 

In analyzing the behavior of the solution to equation (\ref{gravaccretion}), it is instructive to begin by neglecting the term associated with gravitational contraction. Upon doing so, we obtain the following characteristic timescale for accretion-driven spin-up:
\be
\tau_{\rm{accr}} = I (1 + \Gamma) \frac{\langle a_{\rm in}\rangle}{R_\star^0} \frac{M_\star}{\mdot} \sim 40-400 \rm{Myr},
\label{tauaccr}
\ee
where the quoted values correspond to the oft-quoted accretion rate of $\mdot\sim10^{-8} M_\odot$ yr$^{-1}$ and $R_\star^0/ \langle a_{\rm in}\rangle \sim 1/2$. Note that $\tau_{\rm{accr}}$ is substantially longer than a typical disk lifetime. This estimate severely contradicts the notion that accretion of disk material can act to accelerate a star to a substantial fraction of its breakup velocity in a few Myr \citep{bouvier2013}. Indeed, the discrepancy arises from the fact that the usual expression for specific angular momentum accretion rate $\dot{j} = a_{\rm{in}}^2 n_{\rm{in}}$ (see e.g. \citet{1996MNRAS.280..458A}) over-estimates equation (\ref{jspecific}) by a factor of $(\langle a_{\rm in} \rangle / R_\star)^3 (1+\Gamma) \gtrsim 10-100$ because it does not properly account for the morphology of the field.

In light of this argument, we expect that the accretion term in equation (\ref{gravaccretion}) will only result in a small correction to the solution (\ref{gravspinup}). For simplicity, let us assume that $\langle a_{\rm in} \rangle = \mathcal{X} R_\star$, where $ \mathcal{X} \sim 2$ is a constant. Under this prescription, equation (\ref{gravaccretion}) is satisfied by the solution
\begin{eqnarray}
\omega_\star &=& \omega_\star^0 \left(1+\frac{12 \pi (R_\star^0)^3 \sigma T_{\rm{eff}}^4}{b GM_\star^2}t \right)^{2/3} \nonumber \\
&\times& \exp \left(\frac{1}{\mathcal{X} (1+\Gamma)} \frac{\mdisk^0}{M_{\star}}\frac{t}{t+\tau}\right).
\label{gravacrspinup}
\end{eqnarray}

Analyzing the solution in the limit $t \gg \tau$ (i.e. when the
disk has fully dissipated), we can determine the net result of angular
momentum accretion: For an initial disk to star mass ratio of
$\mdisk/M_\star$ = 0.10, and for stars dominated by dipole fields
($\Gamma$ = 0), the rotation rate exceeds the value obtained solely from gravitational contraction by about 5 percent; in contrast, for stars dominated by octupole fields, the rotation rate increases by only about 0.5 percent. The effects of angular momentum accretion are thus small for
dipole systems and negligible for octupole systems.

\subsection{A Comment on Stellar Winds}

Much like accretion of disk material can add angular momentum to the central star, mass carried away by stellar winds drains angular momentum away from the star. Various models for wind-forced rotational evolution are available in the literature. Notably, a semi-analytical treatment of wind-driven angular momentum loss rate has been formulated by \citet{1988ApJ...333..236K}. Since its inception, this prescription has been improved, in part with the aid of additional insight gleamed from numerical simulations (see e.g. \citet{1995ApJ...441..876C, 2012ApJ...745..101M} and the references therein). However, even without actual implementation of the developed models, an upper bound on the angular momentum loss rate can be estimated.

Similarly to the accretionary flow, stellar winds adopt a magnetohydrodynamic force-free profile and travel along magnetic field lines. However, unlike accretion, which takes place along closed field lines, wind must escape the stellar surface along open field lines. An examination of Figure (\ref{setup}) immediately suggests that the polar angles of all field lines along which magnetized wind flows are smaller than that of the critical field line associated with accretion. Aside from this difference, the functional form of wind-driven angular momentum transport is akin to that of the accretionary flow. Consequently, under the disk-locked condition their ratio satisfies:
\begin{equation}
- \left[\frac{d J}{ dt} \right]_{\rm{wind}} \left[\frac{d J}{ dt} \right]_{\rm{accr}}^{-1} \lesssim \frac{\dot{M}_{\rm{wind}}}{\dot{M}_{\rm{accr}}}.
\end{equation}

Although not perfectly constrained, the typically invoked mass-loss rates due to stellar winds are about an order of magnitude smaller than the typical accretion rates\footnote{This notion is expected, since magnetized winds during the PMS stage of stellar evolution are thought to be accretion powered, which implies $\dot{M}_{\rm{wind}}/\dot{M}_{\rm{accr}} < 1$.} i.e. $\dot{M}_{\rm{wind}} \sim 10^{-9} M_\odot$ yr$^{-1}$ \citep{2008ApJ...678.1109M}. As a result, stellar spin-down due angular momentum loss by winds is unlikely to overcome the spin-up that originates from accretion\footnote{Naturally, an exception to this statement arises if the accretionary flow is quenched by an exclusively open field geometry in the vicinity of $a_{\rm{in}}$.}. Furthermore, recall that we have argued that accretion generally constitutes only a small enhancement to the spin-up associated with gravitational contraction. It is therefore likely that the consequences of stellar wind on rotational evolution can be safely neglected during the disk-bearing phase of the stellar lifetime.

\subsection{Magnetic Star-Disk Coupling}
The final effect we shall consider here is magnetic braking due to star-disk coupling. The understanding of the physical processes inherent to magnetic star-disk interactions has evolved considerably since the original proposal of \citet{konigl1991}. However, a detailed description of the model is still unsettled and remains an active area of research \citep{1992MNRAS.259P..23L,1996MNRAS.280..458A,2005MNRAS.356..167M,subu,zanni2009}. Here, we shall follow the more or less conventional treatment presented by \citet{2005MNRAS.356..167M}, while keeping in mind that various alternatives to this picture exist.

The net magnetic torque exerted on the star by the disk is given by the sum of the off-diagonal components of the Maxwell stress tensor, evaluated at $\theta = \pi/2$ \citep{1992MNRAS.259P..23L}:
\begin{eqnarray}
\left[\frac{d J}{ dt} \right]_{\rm{mag}} &=& -\frac{4 \pi}{\mu_0} \int_{\hat{a}_{\rm{in}}}^{\hat{a}_{\rm{out}}} B_{\theta} B_{\varphi} a^2 da \nonumber \\ 
&=& -\frac{4 \pi}{\mu_0} \int_{\hat{a}_{\rm{in}}}^{\hat{a}_{\rm{out}}} \gamma \frac{\mathcal{M}^2}{a^4} da,
\label{magtorque}
\end{eqnarray}
where $\hat{a}_{\rm{in}}$ and $\hat{a}_{\rm{out}}$ represent the inner and outer radii of the magnetically connected region of the disk (i.e. the annulus within which closed magnetic field lines thread the disk) while $\gamma = B_{\varphi}/B_{\theta}$ parameterizes the extent to which field lines are azimuthally twisted by the disk ($\gamma$ is also referred to as the magnetic pitch angle). The sign of the torque is determined by the differential rotation between the disk and the star: if the disk mean motion is greater than the stellar rotation rate, $[dJ/dt]_{\rm{mag}}$ is positive, whereas the opposite is true if the disk material trails the unperturbed rotation rate of the field.

The actual magnitude of the azimuthal twist, $\gamma$, depends on the rate at which the field slips back through the disk diffusively. Accordingly, following \citet{2005MNRAS.356..167M}, we define a dimensionless diffusivity parameter
\begin{equation}
\beta = \frac{\bar{\alpha}}{\mathcal{P}_{\rm{m}}} \frac{h}{a},
\end{equation}
where $\bar{\alpha}$ is the disk viscosity parameter \citep{1973A&A....24..337S}, $\mathcal{P}_{\rm{m}}$ is the magnetic Prandtl number, and $h$ is the scale-height of the disk. As argued by \citet{2005MNRAS.356..167M}, any reasonable choice for the diffusivity parameter is bounded from above by $\beta\lesssim10^{-2}$, an estimate we adopt here.   

In terms of $\beta$, the steady-state twist is given by \citep{2002ApJ...565.1191U}:
\begin{equation}
\gamma = \frac{(a/a_{\rm{co}})^{3/2}-1}{\beta}.
\label{twist}
\end{equation}
It is important to understand that magnetic field lines cannot stretch indefinitely. As shown by \citet{2002ApJ...565.1191U}, beyond a critical twist of order unity, the topology of the magnetic field transitions from closed to open field lines. Thus, equation (\ref{twist}) implicitly defines critical values $\hat{a}_{\rm{in}}$ and $\hat{a}_{\rm{out}}$ as the roots when $|\gamma| = 1$. Clearly, for our adopted value of $\beta$, the magnetically connected region within the disk is quite narrow, with the transition radii not deviating away from the corotation radius by more than $\sim 1\%$.

In light of the dependence of the twist on the distance away from the corotation radius (equation \ref{twist}), it is apparent that the magnitude of the total magnetic torque (equation \ref{magtorque}) is predominately controlled by the exact value of $a_{\rm{in}}$. That is, if the disk is truncated beyond the outer field transition radius ($\hat{a}_{\rm{out}} < a_{\rm{in}}$), then the magnitude of the magnetic torque is null\footnote{Note that in this case, magnetically controlled accretion would also cease \citep{2005MNRAS.356..167M}.}. If the disk is truncated interior to the inner transition radius ($a_{\rm{in}} < \hat{a}_{\rm{in}}$), then the spin-down torques which arise from the part of the disk beyond the corotation radius are to first order cancelled out by the spin-up torques that arise from the part of the disk interior to the corotation radius. Consequently, in this case, it can be expected that $[dJ/dt]_{\rm{mag}} \propto \beta^2$. Magnetic braking is maximized if the disk truncation radius corresponds exactly to the corotation radius i.e. the disk-locked condition ($\hat{a}_{in} = a_{in}$). In this case, we expect that $[dJ/dt]_{\rm{mag}} \propto \beta$. Note that this picture is considerably less optimistic than the one envisioned by \citet{1996MNRAS.280..458A}, where the field was taken to couple to the entire disk. 

The rates of angular momentum transport in the three regimes are summarized as follows:
\begin{equation}
    \left[\frac{dJ}{dt}\right]_{\rm{mag}} = \left\{
       \begin{array}{l}
       \displaystyle 0 \ \ \ \ \ \ \ \ \ \ \ \ \ \ \ \ \ \ \ \ \ \ \ \ \ \ (a_{\rm{in}} > \hat{a}_{\rm{out}}) \\ \\
       \displaystyle - \frac{4 \pi}{3\mu_0}\frac{ \beta \mathcal{M}^2}{a_{\rm{co}} (1+\beta)^2}  \ \ \ \ \ (a_{\rm{in}} = a_{\rm{co}})  \\ \\
       \displaystyle - \frac{16 \pi}{3\mu_0}\frac{ \beta^2 \mathcal{M}^2}{a_{\rm{co}} (1-\beta^2)^2}  \ \ \ \ (a_{\rm{in}} < \hat{a}_{\rm{in}})
    \end{array}   \right.
\label{magsummary}
\end{equation}
Similarly, we can evaluate the corresponding spin-down timescales:
\begin{equation}
    \tau_{\rm{mag}} = \left\{
       \begin{array}{l}
       \displaystyle \infty  \\ \\
       \displaystyle \frac{3 I}{4 \pi \omega_\star^0}\frac{(1+\beta)^2}{\beta} \left(\frac{G M^2 R^2 \mu_0}{\mu^2} \right) \sim 3 \rm{Myr} \\ \\
       \displaystyle \frac{3 I}{16 \pi \omega_\star^0}\frac{(1-\beta^2)^2}{\beta^2} \left(\frac{G M^2 R^2 \mu_0}{\mu^2} \right) \sim 100 \rm{Myr},
           \end{array}   \right.
\label{magtimes}
\end{equation}
where the vertical ordering is identical to equations (\ref{magsummary}). It is noteworthy that the terms inside the parentheses in equations (\ref{magtimes}) approximately correspond to the ratio of hydrostatic pressure in the stellar interior and magnetic pressure at the stellar surface.

From these estimates, we can expect that for values of $a_{\rm{in}}$ chosen in a non-pathological manner, magnetic braking may be competitive with spin-up due to accretion, but not with spin-up due to gravitational contraction. An exception to this statement arises when the disk-locked condition is imposed (albeit in an ad-hoc way), and the magnetic braking timescale approaches the same order of magnitude as the Kelvin-Helmholtz time. Although the disk-locked condition itself is an idealization, due to the episodic nature of accretion, it may be argued that $a_{\rm{in}}$ periodically sweeps across $a_{\rm{co}}$, yielding brief (but possibly frequent) spurts of efficient magnetic braking. Consequently, the extent to which angular momentum exchange via magnetic disk-star coupling is responsible for the observed rotational distribution of T-Tauri stars may be subject to the detailed nature of time-variability inherent to accretion. A thorough exploration of this issue is beyond the scope of this paper.

\subsection{Parameterized Rotational Evolution}

With the relevance of the individual angular momentum transfer mechanisms identified, we shall now perform a few exemplary numerical calculations of the post-embedded rotational evolution of a Sun-like star during its disk-bearing phase. Specifically, we integrate a variant of equation (\ref{gravaccretion}) including the effects of accretion, time evolution of the stellar radius, and external magnetic torques given by equations (\ref{magsummary}). Stellar and disk parameters are taken as before and the magnetic moment is assumed to correspond to a $1.5$ kGauss surface field at $t=0$ ($R_{\star} = R_{\star}^0$). As initial stellar rotation rates, following \citet{gallet}, we adopt $P_\star^0 = 2, 7$, and $10$ days, corresponding to characteristic fast, median and slow rotators, as dictated by the observed distribution. 

\begin{figure}
\includegraphics[width=1\columnwidth]{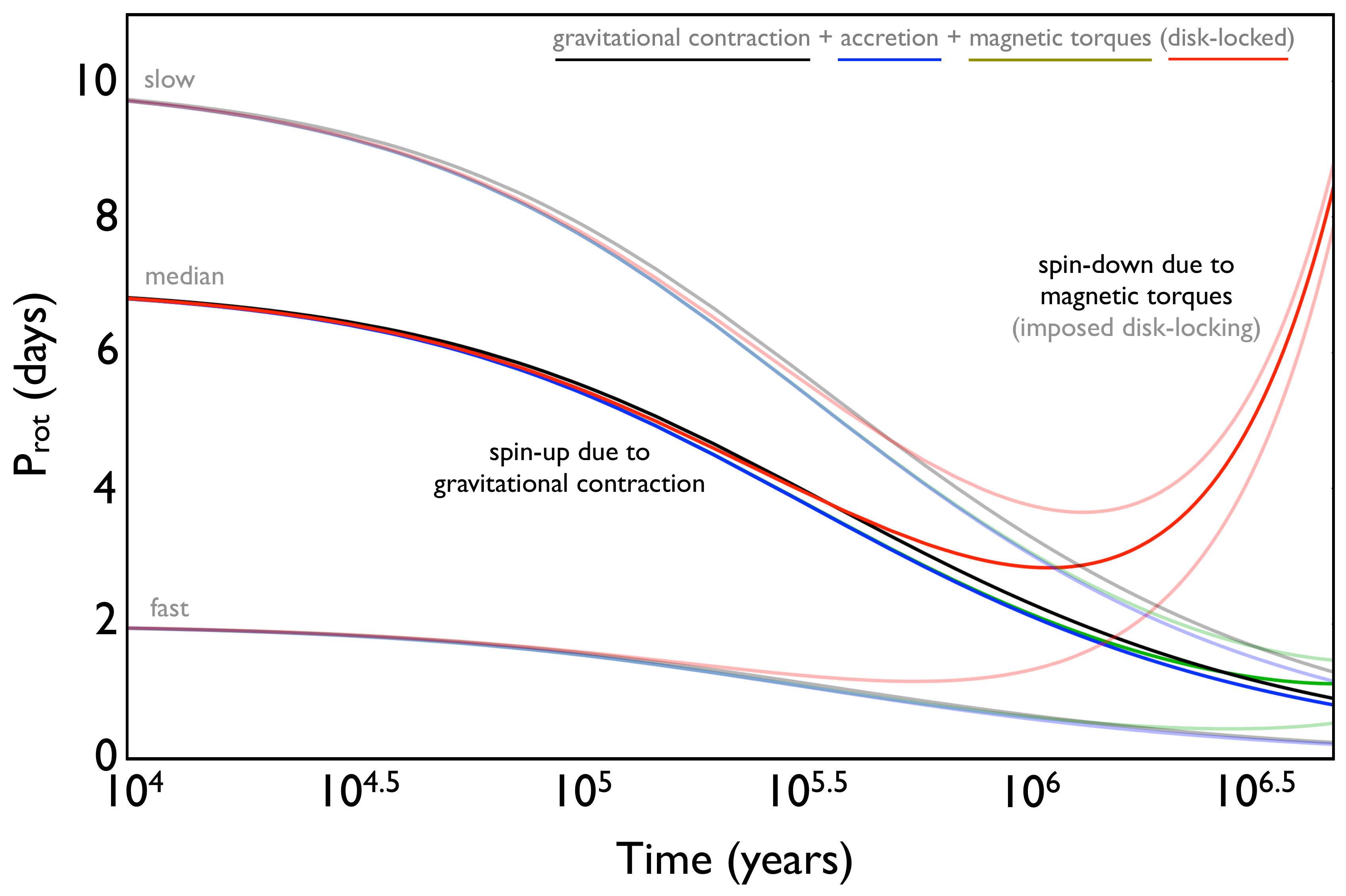}
\caption{Rotational evolution of fully convective ($n=3/2$) stars 
including stellar contraction, accretion of angular momentum
from a disk, as well as magnetic braking. The time evolution of the stellar period is shown for
three different initial conditions: $P_\star^0 = 2, 7$, and $10$ days, corresponding to characteristic fast, median and slow rotators, as dictated by the observed distribution. The final rotational distribution of PMS stars appears to be dictated predominantly by the extent to which the disk truncation radius corresponds to the stellar corotation radius, as this is the primary parameter that determines the effectiveness of magnetic braking. If the truncation radius significantly deviates away from the corotation radius, gravitational contraction dominates the rotational evolution, leading to stellar spin-up.} 
\label{spin}
\end{figure}

The evolutionary tracks are shown in Figure \ref{spin}. For each initial condition, four solutions are presented. Black lines denote isolated spin up due to gravitational contraction (these solutions also correspond to accreting stars, whose fields are dominated by the octupole component i.e. $\Gamma = 10$). Blue lines are solutions that account for accretion in a dipolar-dominated system ($\Gamma = 0$). Solutions that additionally account for magnetic braking in the case of $a_{\rm{in}}<\hat{a}_{\rm{in}}$ are shown with green lines, while red lines correspond to solutions that adopt a magnetic torque with the disk-locked condition. 

As expected, the spin-up due to the reduction in the moment of inertia associated with gravitational contraction dominates the early stages of rotational evolution. Although visible, the effects of accretion and magnetic braking (without assuming the disk-locking condition) clearly contribute only small corrections to the isolated solution. However, under the assumption of disk-locking, magnetic torques dominate the later stages of evolution and effectively de-spin the stars to rotational periods of order the median observed value \citep{affer}. It is noteworthy that the dominance of magnetic torques increases with time. This is a direct consequence of gravitational contraction: as the stellar radius decreases, so does the ratio of interior hydrostatic to surface magnetic pressures, facilitating more efficient braking. 

It is interesting to note that whether the disk-locking condition is imposed on the magnetic torques or not, the variance of the final (evolved) rotation rates is diminished compared to the spread in initial conditions. Furthermore, the rotation rates corresponding to isolated gravitational contraction and those corresponding to magnetically spun-down stars appear to approximately bracket the observed distribution. Thus, if our results are taken at face-value, it may be argued that the presence of both, slow and fast rotators in the observational sample is primarily a consequence of the magnetic torque's sensitive dependence on the truncation radius, leading to variable coupling patterns among different stars. Notice also that the characteristic timescale for the evolution of the stellar spin is of order a few Myr. This suggests the timescales for rotational evolution (Figure \ref{spin}), planet-formation \citep{pollack1996}, and disk removal (e.g., \citealt{jesus}) are all roughly comparable.

Despite the apparent hits of agreement between the above calculations and the observed distribution, it should be kept in mind that our treatment contains a
number of approximations. Most notably, our calculation of magnetic braking is idealized and the extent to which the analogy between the corotation radius and the truncation radius is relevant, remains elusive. As already discussed above, this issue directly ties into the fact that our treatment uses time-averaged quantities to model the accretion flow, which is
fundamentally episodic. This means that the parameter values that should be used are somewhat uncertain and will vary from system to system. Secular changes in the controlling quantities also cannot be ruled out. A much less dominant, but nevertheless important point is that in our calculations, the amount of accreted angular momentum is overestimated (by a small amount)
because we assume that all of the disk material lands on the star; in
actuality, somewhere between a third and a tenth of the disk mass is
carried away by winds that are launched in conjunction with the
accretion flow \citep{shu1994}, while additional stellar braking is
accomplished through stellar winds (see Figure \ref{setup}). Finally, the results shown here use
the dimensionless moment of inertia $I=0.20$ appropriate for a fully
convective star; the moment of inertia could be smaller if the star
develops a radiative core or if only the outer layers are rotating
(and smaller $I$ would amplify the effects of angular momentum
accretion).

In spite of the uncertainties, the characteristic behavior depicted in Figure (\ref{spin}) suggests the following general features for the time dependence of stellar spin. During the early epochs of classical star-disk evolution, gravitational contraction unambiguously leads to stellar spin-up. Subsequently, stars are spun-down (to a variable extent) by torques associated with magnetic star-disk coupling. The final distribution of rotation rates obtained within the context of our calculations appear to be largely independent of initial conditions. This means that even if stars typically emerge from the embedded, protostellar stage as fast rotators, we expect the combined effects of disk-locking and winds/outflows to successfully brake stellar rotation from near-breakup rates to periods of order $P_0\sim7$ days \citep{shu1994,herb2002,herb2007} during the disk-bearing phase. Although the tracks in Figure \ref{spin} are plotted with a single set of assumed controlling parameters, it is implicitly understood that the range of such parameters can be large, and the substantial breadth of the PMS stellar rotational distribution appears to be consistent with the calculations presented above.  Moreover, additional modulation due to
the accretion of disk angular momentum and perhaps other torques will surely act to further homogenize the distribution. 

\section{Dynamics of the Stellar Spin Axis}
\label{sec:dynamics} 

Having characterized PMS stellar rotational evolution due to gravitational contraction, accretion
of disk material, and magnetic braking in the previous section, we now turn our attention to
the orientation of stellar spin axes with respect to planetary
orbits. Specifically, the aim of the following calculation is to
construct a simple quantitative model for the spin-axis dynamics of a
rotating star, forced by gravitational interactions as well as
magnetically controlled accretion arising from a azimuthally-symmetric
circumstellar disk.

Generally, it is expected that the precession timescale of the stellar
pole will greatly exceed both the stellar rotation period and the
orbital period of fluid parcels within the disk. As a result, here we
can work within the context of orbit-averaged perturbation theory
\citep{md2000}. To the extent that a rotating star can be represented
as an oblate spheroid, the dynamics of its spin-pole can be
approximated as those of a point mass surrounded by an orbiting
ring with mass 
\begin{equation}
\tilde{m} = \left[ 
\frac{3 M_{\star}^2 \omega^2 R_{\star}^3 I^4}{4 \mathcal{G} k_2 } 
\right]^{1/3},
\end{equation}
and semi-major axis
\begin{equation}
\tilde{a} = \left[
\frac{16 \omega^2 k_2^2 R_{\star}^6}{9 I^2 \mathcal{G} M_{\star}} 
\right]^{1/3}.
\label{astaryo}
\end{equation} 
To avoid confusion, we adopt a notation where the quantities
corresponding to the above-defined ring are marked by tildas, while
the analogous quantities referring to the disk are marked by primes.

Taking into account the fact that the star is orbited by a massive
proto-planetary disk, the stellar ring will not remain
stationary \citep{bt1987}. Specifically, its longitude of ascending
node will recess. Away from elliptical fixed points in phase space, such a recession
may give rise to associated changes in the inclination. Non-trivial
evolution of the stellar spin-pole is further ensured if the orbital
properties of the disk evolve in time. To a satisfactory
approximation, the dynamical evolution of the stellar spin-axis can be
calculated within the framework of a modified Laplace-Lagrange secular
theory (see, e.g., \citealt{md2000,morbybook}).

It can be trivially shown that angular momenta of proto-planetary
disks (and by extension, planetary systems) greatly exceed that of
even the most rapidly rotating stars. As a result, it is sensible to
calculate the evolution of the stellar spin axis within the framework
of a restricted secular treatment. In other words, we can neglect the
back-reaction of the changes in the stellar spin axis on the
inclination dynamics of the disk. Provided that self-gravity of the
disk is typically sufficiently strong to ensure that the disk behaves
as an effectively rigid body (see \citealt{batygin2011,morby2012}),
the extent of warping within the disk should be negligible, and this 
restricted assumption is likely to be well satisfied for any relevant
choice of parameters.

To begin with, consider the dynamical evolution of the stellar spin
pole due to a massive ring of infinitesimal radial extent, $da'$. As a
starting approximation, we consider the physical parameters of the
star and the disk (i.e., the stellar radius, rotation rate, moment of
inertia, disk mass, etc.) to be constant. This assumption will be
lifted later. To leading order in inclination, $i$, the Hamiltonian
that governs the dynamics of the stellar rotation pole \citep{md2000} 
is 
\begin{eqnarray}
d\mathcal{H} &=& - \frac { \mathcal{G} \tilde{m} dm' }{a'} 
\bigg[ f \sin^2\left( \frac{\tilde{i}}{2} \right) \nonumber \\
&-& 2 f  \sin \left( \frac{\tilde{i}}{2} \right)  
\sin \left( \frac{i'}{2} \right) 
\cos \left(\tilde{\Omega} - \Omega' \right)  \bigg] \,.
\end{eqnarray} 
Here, the quantity $dm' = (2 \pi \Sigma a' da')$ is the mass of the
ring (where $\Sigma \propto r^{-1}$ is the disk surface density),
$\Omega$ is the longitude of ascending node, while $f$ is a constant
that depends only on the semi-major axes $a$. 

It is important to understand that this Hamiltonian only provides a
leading order approximation to the true dynamics of the system. The
solution becomes increasingly imprecise as the mutual inclination
between the stellar spin-axis and the orbital plane of the disk is
increased. As a result, in the vicinity of linear secular resonance,
where the characteristic precession timescale of the star matches the
torquing timescale of the disk, any solution obtained from the above
Hamiltonian will grossly over-estimate the amplitude of
oscillations\footnote{In fact, at exact linear resonance, the solution
  encounters an unphysical singularity that cannot be removed by a change of
  variables.}. That said, away from the near-resonant domain, this
flavor of secular theory has been shown to be a surprisingly good
approximation to higher-order secular solutions\footnote{Although, we note that such correspondence is not assured generally.} \citep{vangreenberg}.
 
As stated above, it is sensible to assume that the inner edge of the
disk, $a'_{\rm{in}}$, is truncated at or near the stellar co-rotation
radius, meaning $n'_{\rm{in}}\simeq \omega_\star$. With this
constraint in mind, it can be easily checked that for any reasonable
set of parameters, $\alpha \equiv \tilde{a}/a'_{\rm{in}} \ll 1$. 
Taking advantage of the smallness of $\alpha$, the expression for $f$
can be simplified considerably. More specifically, after expressing
the Laplace coefficient of the first kind $b_{3/2}^{(1)}(\alpha)$
(see Chapter 7 of \citealt{md2000}) as a hypergeometric series, we can
make the approximation $b_{3/2}^{(1)}(\alpha) \simeq 3 \alpha$. 
As a result, the coefficient $f$ can be expressed in the form 
\begin{equation}
f =  - \frac{3}{2} \left( \frac{\tilde{a}}{a'} \right)^2.
\end{equation}

Because Keplerian orbital elements do not constitute a canonically
conjugated set, we must change variables before proceeding further. 
Specifically, let us convert to modified Poincar\'e
action-angle coordinates: 
\begin{eqnarray}
\label{Poincare}
\tilde{Z} &=& (1 - \cos(\tilde{i})) \ \ \ \ \ \ \ \ \ \ 
\tilde{z} = - \tilde{\Omega} \nonumber \\
Z' &=& (1 - \cos(i')) \ \ \ \ \ \ \ \ \ z' = - \Omega' .
\end{eqnarray}
Note that compared with the usual form of Poincar\'e coordinates (see
Chapter 2 of \citealt{md2000}), the above actions have been
reduced by a factor of $\Lambda = m \sqrt{GM_{\star} a}$. In order to 
consider the above variables as canonical (keeping in mind that
($Z',z'$) are pre-determined functions of time), we must scale the
Hamiltonian by a factor of $\tilde{m} \sqrt{GM_{\star} \tilde{a}}$. 
Utilizing the half-angle formula, we obtain the following expression 
for $d\mathcal{H}$, 
\begin{eqnarray}
\label{Ham2}
d\mathcal{H} = 
\frac{3 \pi \mathcal{G} \Sigma  da'}{2 \sqrt{\mathcal{G} M_{\star} 
\tilde{a}}} 
\left( \frac{\tilde{a}}{a'} \right)^2 \bigg[ \tilde{Z} - 
2 \sqrt{\tilde{Z}} \sqrt{Z'} \cos \left(\tilde{z} - z' \right) \bigg]\,.
\end{eqnarray}

As stated above, the Hamiltonian (\ref{Ham2}) dictates the dynamical
evolution of the stellar spin-axis due to an infinitesimal ring of
mass. In order to obtain a Hamiltonian that accounts for the forcing
that arises from the entire disk, we must integrate $d\mathcal{H}$
radially with respect to $da'$. Noting that the disk mass has the form
\begin{equation}
M_{\rm{disk}} = 2 \pi \int_{a'_{\rm{in}}}^{a'_{\rm{out}}} a' 
\Sigma_{0} \left( \frac{a'_0}{a'} \right) da' \simeq 
2 \pi \Sigma_{0} a'_0 a'_{\rm{out}}\,,
\end{equation}
where $\Sigma_0$ is the surface density at $a'_0$, we define the
characteristic spin-axis precession frequency $\mathcal{F}$ according to 
\begin{eqnarray}
\mathcal{F} &=& \int_{a'_{\rm{in}}}^{a'_{\rm{out}}} 
\frac {3 \pi \mathcal{G} \Sigma_0 a_0 }{2 \sqrt{\mathcal{G} M_{\star} \tilde{a}} a'} 
\left( \frac{\tilde{a}}{a'} \right)^2 da' = \frac{3}{8} \frac{2 \pi \mathcal{G} 
\Sigma_0 a'_0 \tilde{a}^2}{\sqrt{\mathcal{G} M_{\star} \tilde{a}}} \nonumber \\ 
&\times& \left( \frac{1}{(a'_{\rm{in}})^2} -
\frac{1}{(a'_{\rm{out}})^2} \right) \simeq \omega_\star \frac{3}{8}
\left(\frac{\omega_\star}{\tilde{n}} \frac{M_{\rm{disk}}}{M_{\star}}
\frac{a'_{\rm{in}}}{a'_{\rm{out}}} \right).
\label{Fcoeff}
\end{eqnarray}
In this work, we adopt $a'_{\rm{out}} = 50$AU as a characteristic size
of typical protoplanetary disk \citep[e.g.,][]{anderson}. With these 
specifications, the Hamiltonian takes on a simple form 
\begin{eqnarray}
\label{Ham3}
\mathcal{H} = \mathcal{F} \bigg[ \tilde{Z} - 
2 \sqrt{\tilde{Z}} \sqrt{Z'} \cos \left(\tilde{z} - z' \right) \bigg]\,.
\end{eqnarray}
After prescribing a time-dependence for the dynamical state of the
disk, the above Hamiltonian constitutes a system with 1.5 degrees of
freedom. In the regime of interest, however, the system can be reduced
to a single degree of freedom through an appropriate choice of the
coordinate system's orientation.

As shown by \citet{batygin2012}, the longitude of ascending node and
orbital inclination, as measured in a frame coplanar with the
protoplanetary disk at $t=0$, undergoes respectively quasi-linear and
cycloidal oscillations in widely-separated binary stellar systems.
Indeed, such a frame represents the spin-pole orientation of a
non-precessing (i.e., purely spherical) star that is initially in
perfect alignment with the disk. Consequently, the oscillatory
behavior of the disk's inclination was used to argue for primordial
misalignments between hot Jupiter orbits and their host stars.
However, in a reference frame that is coplanar with the orbit of the
stellar companion, the disk inclination (that is initially not null)
is approximately preserved and the recession rate of the disk's
ascending node is constant. In such a reference frame, we may write
$z'=\nu{t}$, where $\nu$ is the disk-torquing frequency and $Z'=Z'_0$.

To make the necessary simplification, we extend the phase space to
four dimensions by adding an action $\mathcal{T}$, conjugated to $t$,
thereby obtaining an autonomous Hamiltonian: 
\begin{eqnarray}
\label{Ham4}
\mathcal{H} = \mathcal{F} \bigg[ \tilde{Z} - 
2 \sqrt{\tilde{Z}} \sqrt{Z'_0} \cos \left(\tilde{z} - \nu t \right)  
\bigg] + \mathcal{T}. \\
\nonumber 
\end{eqnarray}
Next, we perform a canonical change of variables, given by the
following generating function of the second kind:
\begin{equation}
\label{Gen}
G_2 = \left(\tilde{z} - \nu t \right) \Theta + t \Xi.
\end{equation}
Upon application of the transformation equations, we obtain:
\begin{eqnarray}
\label{Newvar}
\tilde{Z} &=& \partial G_2/ \partial \tilde{z} =  \Theta \ \ \ \ \ \ \ \ 
\theta =  \tilde{z} - \nu t \nonumber \\
\mathcal{T} &=& \partial G_2/\partial t =  -\nu \tilde{Z} + \Xi \ \ \ \ \ \  \zeta = t.
\end{eqnarray}
The physical meaning behind the transformation lies in transferring to
a reference frame that is aligned with the orbital plane of the binary
companion but co-precesses with the ascending node of the disk. An
advantage of the new coordinates is that the perfectly aligned state
is always represented by the point ($\Theta, \theta) =
(Z'_0,0)$. Furthermore, after the transformation, the angle $\zeta$ is
absent from $\mathcal{H}$, rendering $\Xi$ a constant of motion that
we subsequently drop. The Hamiltonian is now transformed into a system 
with a single degree of freedom:  
\begin{equation}
\mathcal{H} = \mathcal{F} \bigg[ \Theta - 2 \sqrt{Z'_0} 
\sqrt{\Theta} \cos \left(\theta \right)  \bigg] - \nu \Theta.
\end{equation}

At this point, the Hamiltonian cannot be simplified further. However,
the action-angle coordinates described above posses a coordinate
singularity at $\Theta = 0$, which can be removed by converting to
cartesian coordinates: 
\begin{eqnarray}
\label{cart}
x = \sqrt{2 \Theta } \cos(\theta) 
\qquad {\rm and} \qquad 
y = \sqrt{2 \Theta} \sin(\theta).
\end{eqnarray}
The equations of motion can be written down more succinctly by
defining a single complex variable 
\begin{eqnarray}
\mu = \frac{x + \imath y}{\sqrt{2}} = 
\sqrt{\Theta}\,\mathbf{e}^{\imath \theta}, 
\end{eqnarray}
where $\imath = \sqrt{-1}$. With this definition, 
the Hamiltonian now takes the form 
\begin{equation}
\label{Hcomplx1}
\mathcal{H} = \mathcal{F} \bigg[ \mu \mu^* -  \sqrt{Z'_0} (\mu + \mu^*)  
\bigg] - \nu  \mu \mu^*.
\end{equation}

\begin{figure}
\includegraphics[width=1\columnwidth]{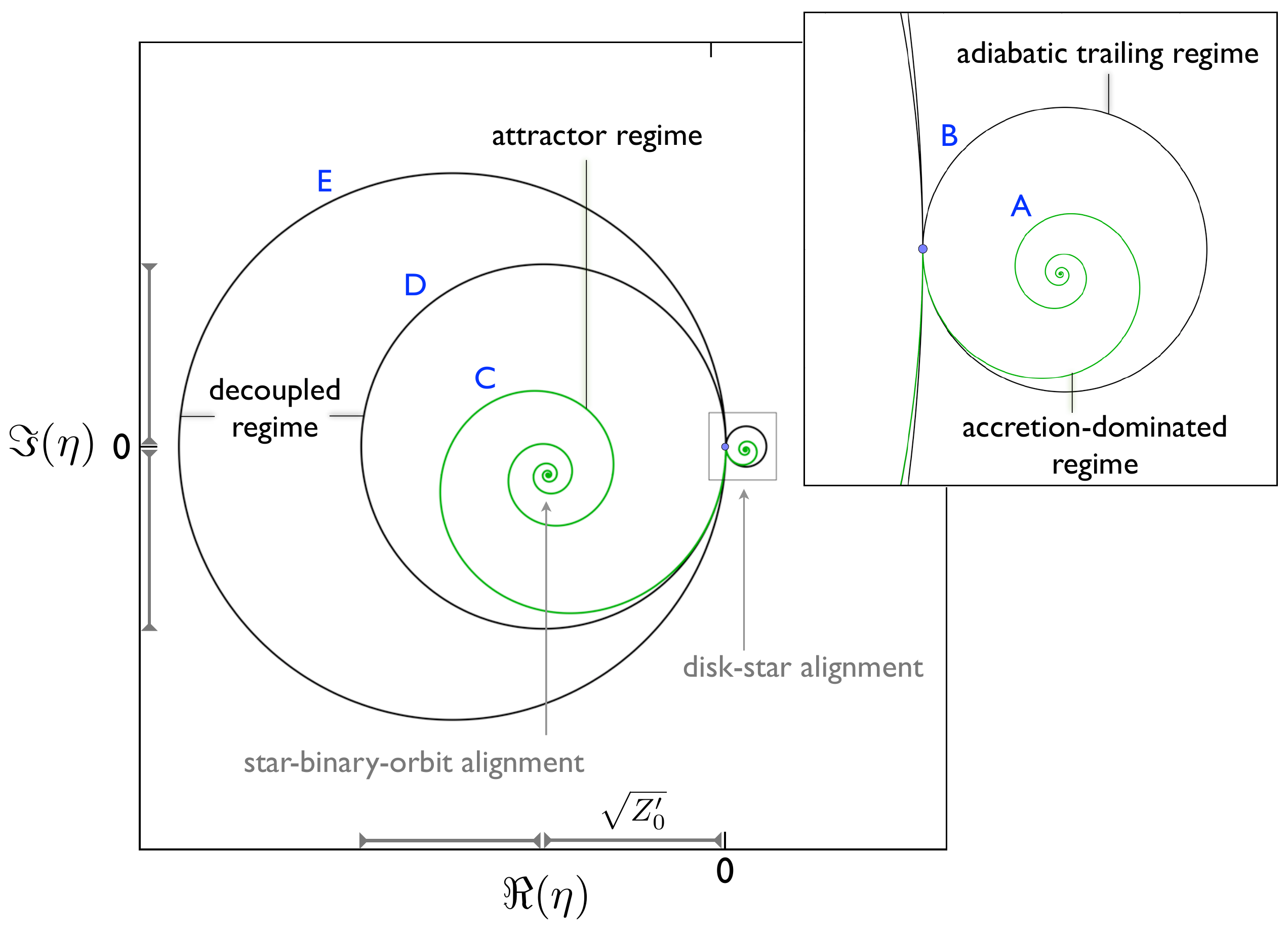}
\caption{Phase-space representation of the characteristic regimes of   
stellar spin axis dynamics (as dictated by equation \ref{etaoft}). 
The green curves represent regimes where accretion plays an
appreciable role, while black curves depict regimes where accretion is
negligible. The origin corresponds to a disk-aligned state of the
star, which also corresponds to the adopted initial condition.
Recalling the variable transformations utilized in the derivation of
equation (\ref{etaoft}), the misalignment angle is obtained by the
distance from the origin to the point on the orbit at which the system
resides at a time $t$, i.e., $\psi = \arccos(1 - \eta \eta^*)$. The
disk inclination, encapsulated into $\sqrt{Z_0'}$, which also sets the
scale of the orbits, is shown with gray lines.}
\label{orbit}
\end{figure}

To motivate one final transformation, let us recall that the
misalignment angle of interest is not measured in a frame aligned with
the binary orbit (where the angle is given by $\mu$), but rather one
associated with the planetary orbit. We thus perform a canonical
translation into a disk-aligned frame, through the definition 
\begin{eqnarray}
\label{translation}
\eta \equiv \mu - \sqrt{Z'_0}. 
\end{eqnarray}
After dropping constant terms, 
this substitution transforms the Hamiltonian into 
\begin{equation}
\label{Hcomplx}
\mathcal{H} = \mathcal{F} \eta \eta^* - \nu \bigg[ \eta \eta^* + 
\sqrt{Z'_0} (\eta + \eta^*)  \bigg].
\end{equation}

The Hamiltonian (\ref{Hcomplx}) governs the conservative
component of disk-star interactions. In certain circumstances, it may be argued that this expression does not
provide a full evolutionary description. For example, in Section
\ref{sec:accretion}, we saw that accretion of disk material onto the
stellar surface and magnetic disk-star coupling can result in limited transfer of angular momentum. Consequently, in systems where the orientations of the stellar spin-axis and the disk's orbital plane are not aligned, dissipative processes can in principle provide corrections to a solution obtained exclusively from equation (\ref{Hcomplx}). 

On a related note, it is important to understand that within the context of stellar spin-axis dynamics, the role of gravitational contraction (which may dominate stellar spin-up) is not analogous to that of accretion or magnetic braking, because the former is a processes that conserves angular momentum (and would operate similarly in absence of the disk). In other words, gravitational contraction itself neither leads to excitation nor damping of mutual star-disk inclination\footnote{In principle, a similar argument applies to stellar winds if the wind profile (i.e. magnetic field geometry) possesses cylindrical symmetry around the stellar spin axis.}. However, there exists an indirect effect associated with gravitational contraction: as stellar radius decreases and the spin is modulated, the coefficient of the Hamiltonian (\ref{Hcomplx}) will also vary in accord with equations (\ref{astaryo}) and (\ref{Fcoeff}). As will be shown below, this plays a central role in determining if the aforementioned encounter with a secular resonance occurs. 

Generally, the details of magnetically-controlled disk-star interactions are rather complex and can only be faithfully captured within the framework of numerical MHD simulations
\citep[see][]{rom02,rom03}. However, as shown in section \ref{sec:accretion}, the characteristic timescale of angular momentum transfer can be estimated analytically. For the purposes of our calculations, it is sensible to adopt such estimates, instead of performing full-fledged numerical MHD simulations. Consequently, here we model the accretion-driven re-alignment of the star as an exponential decay of the mutual inclination angle, and adopt $\tau_{\rm{accr}}$ (as given by equation \ref{tauaccr}) as the \textit{e}-folding time. We note however, that the inclusion of the accretion term is done solely for generality, as its practical effect is negligible.

Evolution of the spin-pole orientation facilitated by magnetic torques is considerably more complicated. Naively, it is tempting to adopt the same functional form for this process as that due to accretion. However, realistically this may not be appropriate, since as shown by \cite{lai2011} (see also \citealt{1998clel.book.....J}) magnetic torques can act to excite, rather than damp mutual star-disk inclination. Furthermore, when acting in conjunction with accretion, magnetic torques may act to evolve the star-disk misalignments towards equilibrium values that sensitively depend on the details of the system at hand. In light of the uncertainties inherent to the physics of magnetic tilting, we shall neglect this process in our model. Indeed, the notion that magnetic tilting appears to be competitive with accretion \citep{lai2011} renders our disregard for its effects reasonable, since the latter operates on timescales somewhat longer than typical lifetimes of protoplanetary disks (see equation \ref{tauaccr}). Still we note that our model may represent a conservative perspective on primordial star-disk misalignments. 

Putting all of the above pieces together, we find that the evolution
of the inclination vector is governed by the differential equation  
\begin{eqnarray}
\label{detadt}
\frac{d \eta}{dt} &=& \bigg[ \imath 
\frac{\partial \mathcal{H}}{\partial \eta^*} \bigg] + 
\bigg[\frac{d \eta}{dt} \bigg]_{\rm{accr}} \nonumber \\
&=& \imath (\mathcal{F} - \nu ) \eta - 
\imath \nu \sqrt{Z'_0} - \frac{\eta}{\tau_{\rm{accr}}}.
\end{eqnarray} 
The solution is characterized by three physical timescales: the
stellar spin-axis precession timescale $2 \pi/\mathcal{F}$, the disk
torquing timescale $2 \pi/\mathcal{\nu}$, and the accretion timescale
$\tau_{\rm{accr}}$. Of course, in reality, all three of these
characteristic timescales are themselves time dependent and evolve
throughout the lifetime of the disk. Specifically, both the spin-axis
precession and accretion timescales are determined by evolving
physical properties of the star and the disk, whereas the
disk-torquing timescale is sensitive to changes in the perturbing
companion's orbit that may arise from the binary's evolution within
its birth cluster or external perturbations from passing stars. In spite of these complications, it is instructive
to examine the limiting cases of the solution analytically.

\begin{figure}
\includegraphics[width=1\columnwidth]{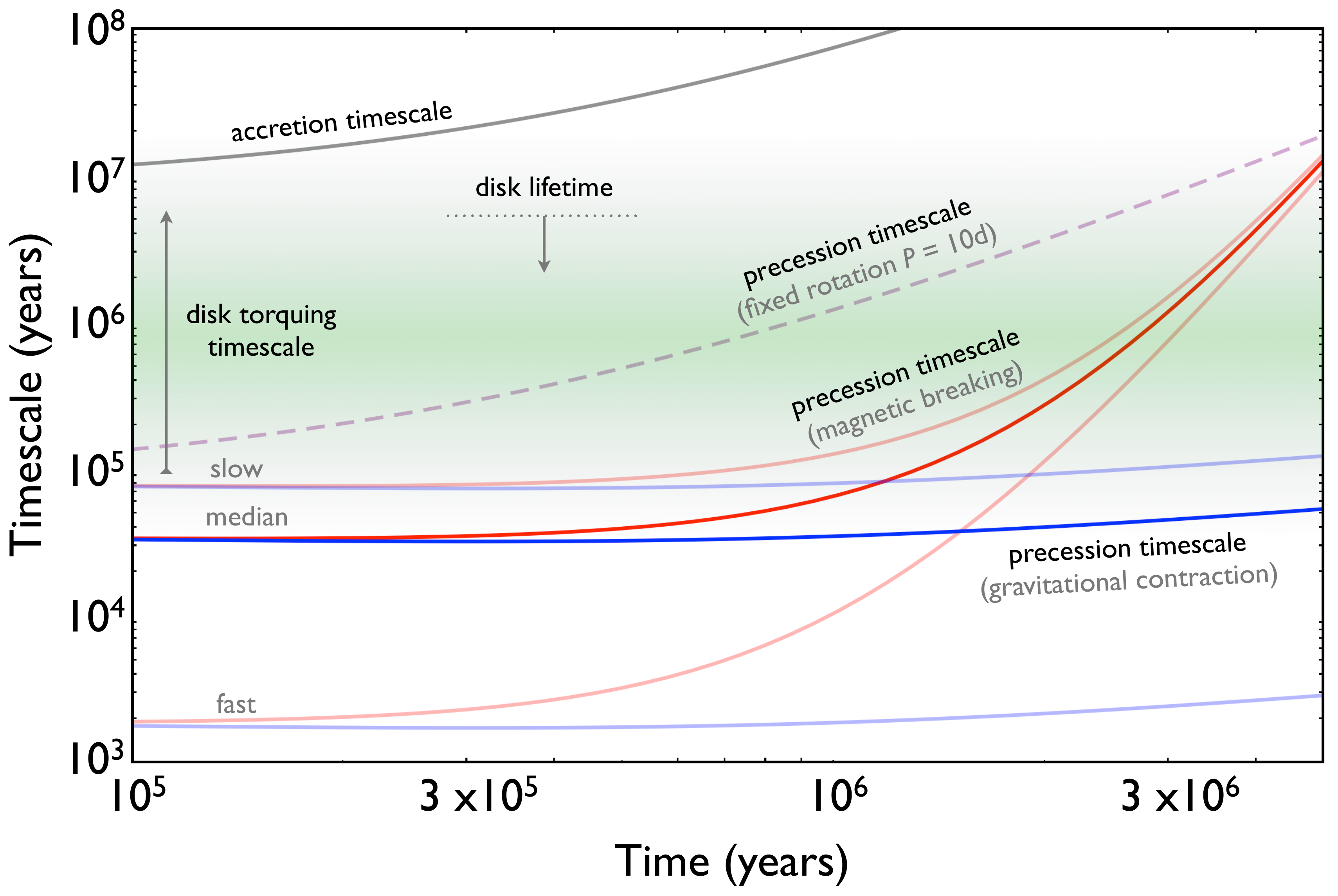}
\caption{Timescales associated with the dynamics of the stellar spin-axis, given here as a function of time. Blue and red curves represent the precession timescale, $2 \pi/\mathcal{F}$, and correspond to the rotational evolution sequences depicted in Figure
\ref{spin}. Accordingly, the blue curves are those where the star is subjected exclusively to gravitational contraction and accretion while the red curves are those where magnetic braking successfully spins the star down to periods of order $\sim 10$ days in $\sim$ 5 Myr. The precession timescale corresponding to a star whose rotational period is held fixed at $P_{\star} = 10$ days is shown as a purple dashed curve. The accretion timescale is shown as a gray curve, whereas typical disk-torquing timescales are represented by a green shade.} 
\label{timescalesyo}
\end{figure}

Assuming a perfectly aligned initial condition $\eta_0 = 0$, as well
as constant values for $\mathcal{F}$, $\nu$, and $\tau_{\rm{accr}}$,
equation (\ref{detadt}) can be integrated to take the form 
\begin{eqnarray}
\label{etaoft}
\eta = \sqrt{Z'_0} \bigg[  \frac{ \mathbf{e}^{\imath(\mathcal{F}-\nu) t 
- t/\tau_{\rm{accr}}} -1 }{1- \imath/(\nu\tau_{\rm{accr}})-F/\nu} \bigg].
\end{eqnarray}
Upon examination of the solution, we can identify four characteristic
regimes, which we list below. For reference, a typical representation
of each regime is depicted in Figure \ref{orbit}. 

\begin{figure*}
\includegraphics[width=1\textwidth]{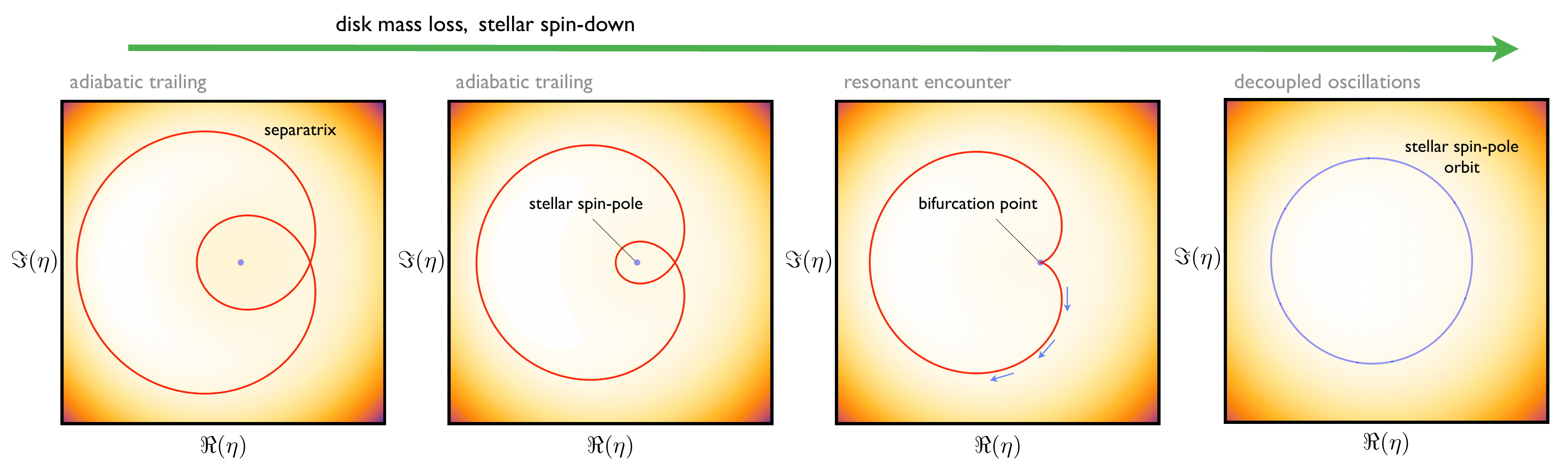}
\caption{A phase-space representation of a typical evolutionary sequence featuring a resonant encounter. As time (represented by a green arrow) marches forward, disk mass loss and magnetic braking of the star cause the parameters of the Hamiltonian to change adiabatically. As the separatrix (shown in red) shrinks, eventually the stable (blue) equilibrium point (on which the stellar spin axis is initially taken to reside), and the unstable equilibrium point (i.e. the x point on the separatrix) of the Hamiltonian join, forcing an excitation of spin-orbit misalignment. The scale of the figure is set in part by the initial disk-binary orbit separation and is purposely not labeled. Background color is representative of level curves of the Hamiltonian.} 
\label{resenc}
\end{figure*}

\textit{Accretion-dominated regime}. Arguably, the simplest dynamical
picture is one where the disk-torquing timescale is exceptionally long, and accretion dominates. This occurs in the limit
$1/(\nu \tau_{\rm{accr}}) \gg (1,\mathcal{F}/\nu$). Accordingly, the
denominator in equation (\ref{etaoft}) retains only the accretion
term, while the exponential in the numerator vanishes over a
comparatively short timescale. Independent of the particular initial
condition, the (diminutive) value of the misalignment angle is
controlled entirely by the smallness of $\nu \tau_{\rm{accr}}$ 
and the solution reduces to the form  
\begin{eqnarray}
\label{etaaccr}
\eta^{(\rm{accr})} \rightarrow \imath \sqrt{Z'_0} 
(\nu\tau_{\rm{accr}}) \simeq 0.
\end{eqnarray}
This regime is characteristic of very young (embedded) systems where the disk is
massive and the accretion rate is maximal, ensuring that any
excitation of the misalignment angle arising from external
perturbations of the proto-planetary disk damps away rapidly. A phase-space
representation of an accretion-dominated solution is shown as a green
spiral, labeled A in the small inset of Figure \ref{orbit}. Here, 
the phase-space trajectory spirals into the origin.  

\textit{Adiabatic trailing regime}. Another extreme parameter regime
that prevents significant excitation of misalignment is one where
efficient disk-star angular momentum transfer is accomplished through
gravitational coupling. For the sake of definitiveness, suppose that
accretion is completely negligible (i.e., $\tau_{\rm{accr}} = \infty$)
but the spin-axis precession timescale is much shorter than the
external torquing timescale, so that $\mathcal{F}/\nu \gg 1$. In this
regime, we obtain an oscillatory solution, where the amplitude is small, 
$|\eta|\sim\nu/\mathcal{F}$, and the solution approaches the form  
\begin{eqnarray}
\label{etaad}
\eta^{(\rm{ad})} \rightarrow \sqrt{Z'_0} \left( \frac{\nu}{\mathcal{F}} \right)
\left[  1 - \mathbf{e}^{\imath\mathcal{F} t } \right] \simeq 0.
\end{eqnarray}
In the inset of Figure \ref{orbit}, this adiabatic-trailing regime is
shown with a black (circular) orbit, labeled B. This type of behavior is
characteristic of systems where the stars are spun up at early times;
the stars can then develop a strong gravitational quadrupole moment,
which facilitates gravitational coupling between the star and the disk. We
note that this regime may also be relevant for systems where
the gravitational torquing of the disk is exceptionally slow (e.g.,
when the disk has a small radius, or the binary companion has a wide
orbit), so that even a modestly deformed star can keep up with the
twist.

\textit{Attractor regime}. Now we consider a system that does not
satisfy either of the above criteria. For simplicity, imagine a
perfectly spherical ($\mathcal{F} = 0$), accreting star, whose
re-alignment timescale is not overwhelmingly longer than the
disk-torquing timescale, i.e., $\nu\tau_{\rm{accr}}\sim 5-10$. In
this case, over a few disk-torquing times, the exponential in equation
(\ref{etaoft}) decays away while the denominator is approximately
unity. The solution in this regime reduces to the form  
\begin{eqnarray}
\label{etaatt}
\eta^{(\rm{att})} \rightarrow - \sqrt{Z'_0}.
\end{eqnarray} 
An orbit of this type is shown in Figure \ref{orbit} as a big green
spiral labeled C. The physical meaning of this solution is most easily
understood by recalling the variable transformation of equation
(\ref{translation}). In the frame aligned with the binary orbit, the
above limit corresponds to a null inclination, $\mu = 0$, which
represents alignment between the angular momentum vectors of the
binary orbit and the stellar spin. Naturally, in this regime, the 
star-disk misalignment angle becomes identically equal to the
misalignment angle between the primordial disk and the binary orbit.
Intuitively, this end-state is sensible because it corresponds to an
equilibrium, reached upon phase-averaging the angular dependence of
accretion from a rapidly torqued disk.

\textit{Decoupled regime}. Finally, we consider the regime where the
timescales associated with precession of the stellar spin axis and
accretion are both much longer than the disk-torquing timescale,
implying that $\nu\tau_{\rm{accr}}\gg1$ and $\mathcal{F}/\nu\ll1$.  
In this case, the denominator in equation (\ref{etaoft}) is (again)
approximately unity, while the accretion term in the exponential can be
dropped, yielding a solution of the form 
\begin{eqnarray}
\label{etadec}
\eta^{(\rm{dec})} \rightarrow \sqrt{Z'_0} 
\left[  \mathbf{e}^{\imath(\mathcal{F}-\nu) t } -1  \right].
\end{eqnarray}
It can be shown that in terms of the original ($\tilde{Z},\tilde{z}$)
variables from equation (\ref{Poincare}), this solution corresponds to
a state where the star remains at its initial condition for all
time. The phase-space trajectories associated with this solution are
shown as black curves in Figure \ref{orbit}, labeled D (with
$\mathcal{F}/\nu = 0$) and E (with $\mathcal{F}/\nu = 1/3$). Indeed,
the non-triviality of the above limit arises entirely from fixing the
inclination and longitude of ascending node of the reference frame to
coincide to that of the disk at all times. As shown by
\citet{batygin2012} through an alternative derivation, in this regime
the disk-star misalignment angle can be represented parametrically by
a cycloid.

As disk and stellar properties evolve with time, the characteristic
regimes of the system will change. Thus, to obtain a quantitatively
accurate solution for the stellar spin-axis dynamics, an explicitly
time-dependent counterpart of the Hamiltonian (\ref{Hcomplx}) must be
considered. Before doing so, however, we note that the overall
behavior of the system can be understood qualitatively by examining
the evolution of the above-mentioned timescales through a typical
lifetime of the disk. Such a comparison is shown in Figure
\ref{timescalesyo}, where precession timescales corresponding to accreting and gravitationally contracting stars are shown in blue while those spun down by disk-locked magnetic torques are shown in red (see Figure \ref{spin}). Additionally, the precession timescale for a star with a fixed rotational period of $P_\star = 10$ days is shown by a purple dashed curve, and the accretion timescale is shown as a gray line. Meanwhile the green shade represents a characteristic range of binary perturbation timescales of interest (see \citealt{batygin2012} for an in-depth discussion).

Upon examining Figure \ref{timescalesyo}, some conclusions about the likely nature of the solution can be drawn. First, the accretion-dominated regime seems to be irrelevant for classical disk-bearing T-Tauri systems, although it is quite possible that this regime is pertinent during the embedded stage of a typical star's evolution. Second, the evolutionary difference between fast and slow rotators is striking: the precession timescale associated with stars that spin-up due to gravitational contraction and accretion remains approximately constant through out the disk's lifetime and consistently straggles behind the average disk-torquing timescale. This means that fast rotators usually reside in the adiabatic trailing regime and misalignments can be expected to be rare. Meanwhile, the precession timescale of stars that successfully undergo magnetic braking increases by many orders of magnitude over a typical disk lifetime. Although such stars also start out adiabatically trailing the disks, the latter stages of their evolution are better characterized by the decoupled regime, where the disk-star inclination can reach substantial values. Note that the evolution of the precession timescale for a star rotating with a fixed period is qualitatively similar to that associated with a star that is magnetically spun-down.

\begin{figure}
\includegraphics[width=1\columnwidth]{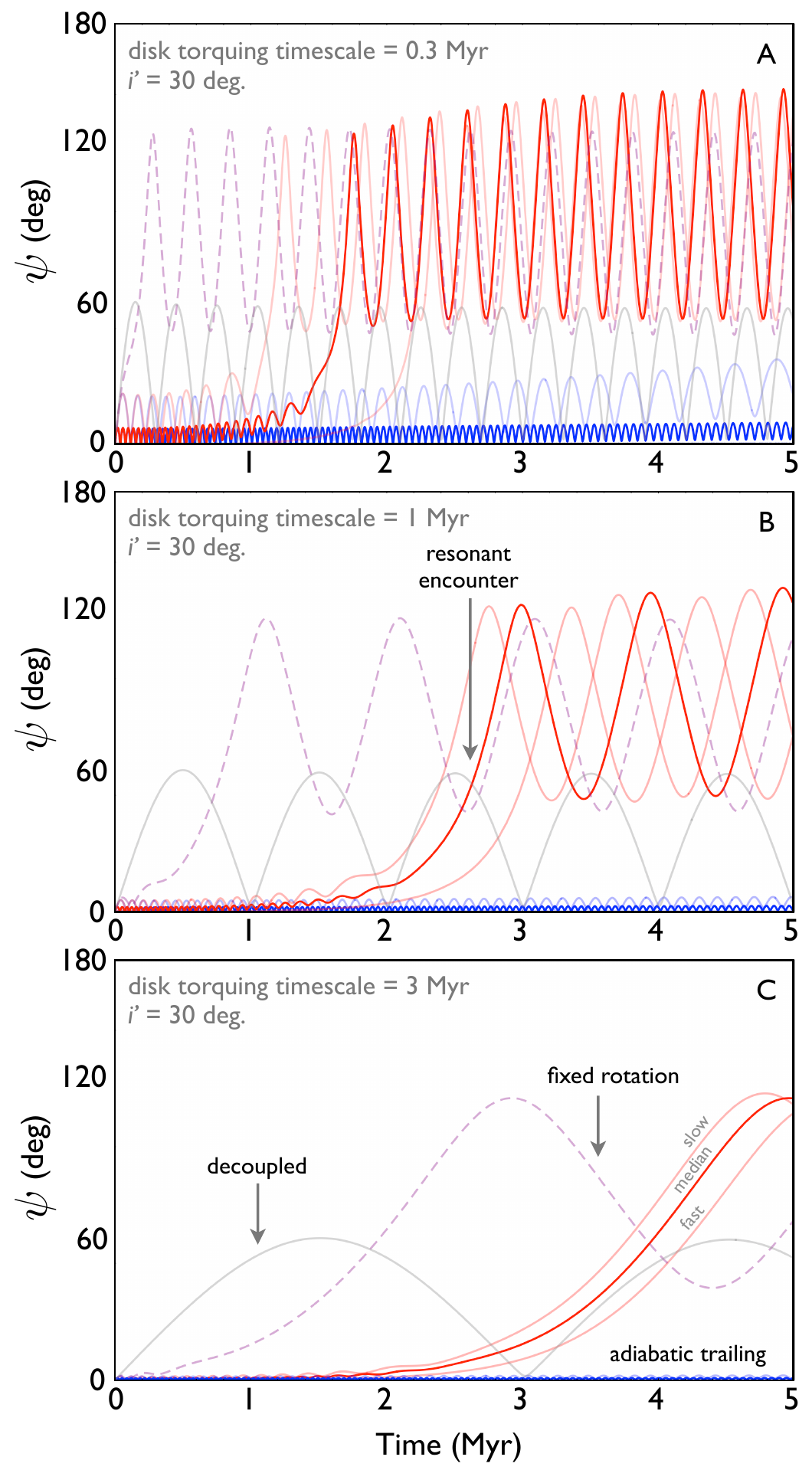}
\caption{Misalignment angle as a function of time, computed within the  
framework of the nonlinear Hamiltonian from equation (\ref{Hnonlin}), 
augmented with a simple prescription for accretion-driven re-alignment
of the star. The utilized color scheme is largely the same as that used in
Figures \ref{spin} and \ref{timescalesyo}. For completeness, a solution with $\mathcal{F} = 0$ is also shown as a gray line. The three panels of the
figure depict solutions with different disk-torquing timescales, as
labeled. The initial disk-binary orbit misalignment is taken to be 
$i' = 30$ degrees across the panels. Note that because the solution is
obtained from a non-linear Hamiltonian, the result will not 
necessarily scale with the disk-binary inclination, as in the case of
the linear solution of equation (\ref{etaoft}).}
\label{psi}
\end{figure}

With the above treatment in mind, it is important to note that as the
characteristic timescales evolve, there are (relatively short-lived)
epochs when the precession and disk-torquing timescales are comparable. Until now, we have
purposely avoided the discussion of such a (resonant) regime, because a commensurability between the frequencies
$\mathcal{F}$ and $\nu$ leads to a singularity in the solution
(\ref{etaoft}). To circumvent this problem, additional terms in the series expansion of
the Hamiltonian must be retained. Following \citet{lithwickwu2012}, we
incorporate an additional non-linear kinetic term, whose coefficient
is coincidentally the same as that of the leading-order terms, $f$, to
leading order in $\alpha$ \citep{md2000}. Working through the
transformations outlined above, the Hamiltonian takes the form 
\begin{equation}
\label{Hnonlin}
\mathcal{H} = \mathcal{F} \bigg[ \eta \eta^* - 
\frac{( \eta \eta^*)^2}{2} \bigg]- \nu 
\bigg[ \eta \eta^* + \sqrt{Z'_0} (\eta + \eta^*)  \bigg],
\end{equation}
where the expansion in inclination is taken relative to the disk's frame.

The Hamiltonian (\ref{Hnonlin}) is a complex representation of the
largely successful second fundamental model for resonance
\citep{henrardlamaitre}. Within the framework of this model, and the
closely-related pendulum model, the theory of resonant encounters is
well-established (for more in-depth treatments, see, e.g.,
\citealt{goldreichpeale,yoder1973,peale1986,henrardbible}), and
qualitatively, the process boils down to the following picture. As
system parameters change, the location of the homoclinic critical
curve (the separatrix) evoloves in phase-space. When the actual
trajectory of a system encounters the separatrix, the system
experiences an impulsive change in action. A schematic phase-space representation of a typical evolutionary sequence, where a resonant encounter occurs is shown in Figure \ref{resenc}.

Figure \ref{psi} shows a collection of solutions obtained by
numerically integrating the equation of motion arising from the
Hamiltonian (\ref{Hnonlin}), augmented with a dissipative term arising
from accretion, as in equation (\ref{detadt}). The color scheme used
here is the same as that in Figures \ref{spin} and \ref{timescalesyo}. Specifically, solutions depicted with red curves are those where spin-down due to magnetic braking is evident and those depicted with blue curves correspond to rapidly rotating stars. The panels A, B and C, depict solutions with the disk torquing
timescale taken to be $\tau_{\rm{torq}}$ = 0.3, 1 and 3 Myr,
respectively. The inclination between the disk and the binary orbit is taken to be $i' = 30$ degrees across the panels. 

Evidently, two regimes characterize the solutions, depending on the rotational evolution of the star. As expected from Figure \ref{timescalesyo}, the spin axes of stars that are spun up due to gravitational contraction and accretion adiabatically trail the disk, leading to consistently low misalignments. Stars that are gradually spun-down by magnetic disk-star coupling encounter a resonance between the spin-axis precession frequency and the disk torquing frequency. As discussed above, this results in an impulsive excitation of mutual inclination, that is followed by oscillations characteristic of the decoupled regime. Similar behavior can be seen for the case where the stellar rotation period is held fixed (shown with a dashed purple line). For completeness, solutions corresponding to a perfectly spherical star such as those considered by \citet{batygin2012}, which reside in the decoupled regime by definition, are also shown with gray curves.

Cumulatively, our calculations show that substantial star-disk inclinations naturally arise for a wide range of binary orbit configurations, a proxy for which is the disk-torquing timescale. However, a controlling feature that determines the feasibility of misalignment excitation is stellar rotational evolution. For disk torquing timescales of order $\sim$ Myr or longer, fast rotators gravitationally lock to the disk inhibiting the onset of misalignment, while median and slow rotators tilt as a result of an encounter with a secular stellar precession-disk torquing resonance. Although the latter holds true for faster disk-torquing timescales, as elucidated by panel A of Figure \ref{psi}, free oscillations of the misalignment angle can be substantial even for rapid rotators. Accordingly, the inherent breadth of the distribution of rotation rates of young stars appears to be an essential feature, required for the reproduction of the observed wide-ranging distribution of Hot Jupiter spin-orbit misalignments. 

\section{Conclusion and Discussion}
\label{sec:conclusion} 

In this paper, we have characterized long-term disk-star interactions, with an eye towards consolidating rotational evolution of young stellar objects and the generation of spin-orbit misalignments into a unified framework. In this initial effort, we have placed significant emphasis on analytic treatment of the physics, yielding an approximate, yet more or less transparent model that captures the basic ingredients of the problem at hand.

 We began by considering the evolution of stellar rotation rates, accounting for gravitational contraction, accretion, and magnetic torques. Our calculations show that depending on the relaxation or imposition of the disk-locking assumption, stars end up with rotation rates that fall in the observed range of $P = 1 - 10$ days \citep{herb2002,littlefair,affer} after several Myr. Stellar spin-down appears to be dominated by magnetic torques, while gravitational contraction dominates stellar spin-up. A careful account for the geometry of the magnetic field suggests that accretion is a sub-dominant process. The parity of gravitational contraction and magnetic braking yields considerable diversity in spin rates within the observed range, where the value for a particular system depends on the exact choice of parameters. The heterogeneity inherent to rotational evolution is further enhanced by the corrections to the spin-up torques introduced by accretion, and its sensitive dependence on the dipole/octupole morphology of the magnetic field.

Although we haven't considered this complication in our calculations, it is noteworthy that the dipole/octupole dichotomy is also likely to add diversity to the evaluation of magnetic torques. As an example, consider a pair of stars with the same surface field strengths, but different field structures. Because an octupole field decays more rapidly with distance, we can expect a weaker magnetic field at the disk coupling point, implying a smaller spin-down torque, compared to the dipole-dominated counterpart \citep{1992MNRAS.259P..23L,2005MNRAS.356..167M}. Further complexity is introduced by the fact that the magnetic poles can be tilted with respect to the rotational poles, and the dipole/octupole orientations can exhibit large inclinations with respect to each-other. All of these considerations may give rise to substantial quantitative changes of the solutions shown in Figure \ref{spin} as well as greater difficulty in trying to ascertain which stars end up with which rotational periods.

In the latter half of the paper, we considered the spin-axis dynamics of rotationally evolving, disk-bearing stars, subject to external torques arising from binary companions. Using orbit-averaged perturbation theory, we demonstrated that large misalignments between disk orbital angular momentum vectors and stellar spin axes are preferentially excited in systems where rotational evolution features a significant spin-down of the host star. Specifically, in nearly all of the solutions we obtained (with the exception of those characterized by rapid disk-torquing timescales), gravitational coupling between the disk and the stellar gravitational quadrupole moment (which arises from rotational deformation) played a dominant role in quenching spin-orbit misalignment at early epochs. However, as the disk mass declined, allowing gravitational coupling to subside, stars whose rotation rates decreased (or were held constant) experienced an impetuous excitation of spin-orbit misalignment as a result of an encounter with a disk-star precession resonance. Such a resonance was never encountered for stars whose rotation rates monotonically increased throughout the disk's lifetime. In either case, star-disk realignment due to accretion was consistently negligible.

It is important to understand that the manner in which the spin-orbit misalignment originates within the calculations presented here is qualitatively different from that envisioned by \citet{batygin2012}. That is, the decoupled regime characteristic of the solutions presented by \citet{batygin2012} is only observed after the initial orbital obliquity is seeded. Indeed, the realization of an impulsive acquisition of spin-orbit misalignment due to a crossing of a secular resonance between the disk-torquing frequency and the stellar precession frequency is the key result of this paper.

The theory for the origin of hot Jupiter spin-orbit misalignments presented here can be confirmed or refuted observationally. The most direct avenue for doing so is a survey aimed at detection and characterization of close-in giant planets around T-Tauri stars. If disk-driven migration (as opposed to dynamical N-body interactions) is responsible for the production of hot Jupiters, the fraction of hot Jupiter-bearing weak-lined T-Tauri stars should be commensurate with (or perhaps greater than) that of field stars i.e. $\gtrsim 1\%$ \citep{2011PASP..123..412W}. Our model would thus predict that among such a sample, planets with high obliquities would preferentially reside in multiple stellar systems and orbit slowly rotating stars. To date, with the exception of the recent work of \citet{2013ApJ...774...53B}, such observations remain scarce.

Another test of our model for spin-orbit misalignment may arise from measurements of the Rossiter-McLaughlin effect in systems of multiple low-mass, multi-transiting planets, many of which have now been discovered by the \textit{Kepler} mission \citep{2013ApJS..204...24B}. Because the calculations presented here do not consider planetary mass as an inherent parameter, the mechanism for the acquisition of orbital obliquity should be oblivious to the specific nature of the planetary system at hand. In fact, for our purposes, even the question of distant (beyond the ice-line) formation followed by extensive disk-driven migration (\citealt{2012ARA&A..50..211K} and the references therein) vs. in-situ formation \citep{2012ApJ...751..158H,2013MNRAS.431.3444C} should not play an appreciable role. Studies of this sort are already underway \citep{2013ApJ...771...11A} although the number of analyzed systems still remains small. Finally, considerable insight into the importance of the mechanism outlined in this paper may be gleamed from a comparison of the theoretically obtained distribution of spin-orbit misalignments and its observed counterpart. Such an analysis will be published as a follow-up study \citep{crida}.

Independent of the aforementioned tests, there exists an additional observational fact, relevant to our model. The current aggregate of data associated with the Rossiter-McLaughlin effect suggests that the existence of this misalignment depends on stellar mass or photospheric temperature $\teff$. More specifically, the misalignments are essentially null for values of $\teff$ below a critical value of  $T_C \approx 6250$ K, and can be arbitrarily large (that is, ranging from prograde and aligned, to retrograde and anti-aligned) for hotter stars. The theoretical basis for this dichotomy remains controversial. \cite{winn2010} and subsequently, \citet{albrecht2012} have argued that the polarity of the observations reflects the changes in efficiency of tidal dissipation with stellar type \citep{lai2011,lai2012}. Specifically, these authors envisioned a scenario where all hot Jupiters arrive onto their orbits with high inclinations, that are subsequently erased in low-mass stars due to efficient tidal dissipation in the convective parts of the envelope. However, this story was recently challenged by \cite{rogerslin2013}, who argued that tidal dissipation preferentially leads to prograde aligned, retrograde aligned, or orthogonal spin-orbit angles, in some contradiction with the observed distribution.

In principle, the mechanism discussed here is consistent with the narrative of \citet{winn2010}. That is, one can envision an initially isotropic distribution of orbital obliquities, that is further shaped by tidal dissipation, within the framework of our mechanism. However, it is likely that an alternative scenario may also be conjured up. Specifically, based on the interdependence of rotational evolution and disk-star interactions, we may hypothesize that the observed mass-dependence inherent to rotational evolution \citep{bouvier2013} may be responsible for the observed dependence of spin-orbit misalignment on $\teff$. A quantitative examination of this speculative idea would require enhanced understanding of PMS star-disk angular momentum transfer.

In light of the arguments presented above, it is clear that in addition to an enhanced aggregate of observations, a meaningful comparison between the data and the model will require further theoretical developments. Some enhancements to the treatment at hand are straightforward. For example, our polytropic structure models can be replaced with state-of-the-art stellar evolution calculations \citep{2013arXiv1301.0319P}. Moreover, our perturbative treatment of gravitational star-disk interactions (that assumes small inclinations) can be easily cast into a more precise form \citep{1966AJ.....71..891C,1991pscn.proc..193H}. On the other hand, the entirety of the disk-locking process is complicated and imperfectly understood (see \citealt{barnes2001,mattpud,long2005,zanni2009}, and references therein). Consequently, the construction of a significantly more detailed iteration of our model may be constrained by an incomplete understanding of magnetic disk-star coupling. In any case, however, the available observational data suggests that our treatment of rotational evolution (although approximate) is plausible and may be sufficient for the purposes at hand.

In conclusion, the model developed in this work opens up a previously unexplored avenue towards understanding star-disk evolution and its consequences in a unified manner. Although a great deal of additional effort, both observational and theoretical, is required to elucidate all of the physics involved, the novel mechanism presented here provides a promising way to understand a number of observed features of star-disk-planet systems.

\medskip 

\textbf{Acknowledgments:} We thank the referee, Scott Gregory, for providing an exceptionally helpful and thorough report which resulted in a substantial improvement of the manuscript. K.B. acknowledges the generous support from the ITC Prize Postdoctoral Fellowship at the Institute for Theory and Computation, Harvard-Smithsonian Center for Astrophysics. F.C.A. acknowledges support from NASA Origins grant NNX11AK87G.

\end{document}